\documentclass[useAMS,usenatbib]{mn2e}

\usepackage{graphicx,bm,amsmath,multirow,subfigure,longtable,float}
\usepackage{todonotes}
\usepackage[T1]{fontenc}
\usepackage{lmodern}
\usepackage{amsfonts}
\usepackage{hyperref}
\usepackage[multidot]{grffile}

\newcommand\red[1]{#1}

\title[Selection biases in $p(z)$ methods]{Selection biases in empirical $p(z)$ methods for weak lensing}
\author[D. Gruen and F. Brimioulle]
{\parbox{\textwidth}{D. Gruen$^{1,2}$\thanks{Einstein Fellow; e-mail:
dgruen@stanford.edu} and F. Brimioulle$^{3}$}\vspace{0.4cm}\\
\parbox{\textwidth}{
$^{1}$SLAC National Accelerator Laboratory, Menlo Park, CA 94025, USA\\
$^{2}$KIPAC, Physics Department, Stanford University, Stanford, CA 94305, USA\\
$^{3}$Max Planck Institute for Extraterrestrial Physics, Giessenbachstrasse, 85748 Garching, Germany\\
}}

\begin{document}

\date{}

\pagerange{\pageref{firstpage}--\pageref{lastpage}} \pubyear{2016}

\maketitle

\label{firstpage}

\begin{abstract}
To measure the mass of foreground objects with weak gravitational lensing, 
one needs to estimate the redshift distribution of lensed background sources. This is commonly 
done in an empirical fashion, i.e.~with a reference sample of galaxies of known spectroscopic redshift, 
matched to the source population. In this work, we develop a simple decision tree 
framework that, under the ideal conditions of a large, purely magnitude-limited reference sample, 
allows an unbiased recovery of the source redshift probability density function $p(z)$, as a function of magnitude and colour. 
We use this framework to quantify biases in empirically estimated $p(z)$ caused by selection effects 
present in realistic reference and weak lensing source catalogues, namely (1) complex selection 
of reference objects by the targeting strategy and success rate of existing spectroscopic surveys and 
(2) selection of background sources by the success of object detection and shape measurement at low signal-to-noise. For intermediate-to-high redshift clusters, and for depths and filter combinations appropriate for ongoing lensing surveys, we find that (1) spectroscopic selection can cause biases above the 10 per cent level, which can be reduced to $\approx5$ per cent by optimal lensing weighting, while (2) selection effects in the shape catalogue bias mass estimates at or below the 2 per cent level. This illustrates the importance of \emph{completeness} of the reference catalogues for empirical redshift estimation.
\end{abstract}


\section{Introduction}

Weak gravitational lensing enables measurements of the mass distribution of foreground structures 
in the Universe by the distortion that their tidal gravitational fields impose on the images 
of background galaxies. The amplitude of the weak lensing shear $\gamma_t$ is proportional to both 
the surface mass density of the structure acting as the lens and the geometry of the observer-lens-source system,
\begin{equation}
\gamma_t\propto D_d\frac{D_{ds}}{D_s}=: D_d\beta \; ,
\end{equation}
where $D_d$, $D_{ds}$ and $D_{s}$ are the angular diameter distances from observer to lens, and from lens and observer to source, respectively \citep[e.g.][their section 3.1.2]{2001PhR...340..291B}, and we have defined $\beta$ as the ratio of the latter.
In order to use lensing to constrain the properties of the lens and the expansion of the Universe, 
one therefore needs to know the redshifts of lensed sources or -- sufficiently, because weak 
lensing is only ever measured on ensembles of sources -- their redshift distribution.

Knowledge of true source redshifts, e.g. from spectroscopic follow-up of each source galaxy, 
would allow for a lensing analysis that is both unbiased (because one could account for the 
geometry of lenses and sources exactly) and statistically optimally powerful (because one 
could select only true background sources and weight each source galaxy according to its 
expected signal in a minimum-variance estimator of mean shear). Unfortunately, however, 
complete, spectroscopic follow-up of large samples of faint galaxies is not feasible with 
present technology. One therefore has to resort to approximate methods based on photometric 
properties of the source galaxy population.

When high quality photometric information is available, redshifts of source galaxies can be 
determined from fitting redshifted template galaxy spectral energy distributions (SEDs) to 
the photometric flux measurements \citep[e.g.][]{1986ApJ...303..154L,1999MNRAS.310..540A,2000ApJ...536..571B,2001defi.conf...96B,2006A&A...457..841I,2008ApJ...686.1503B}. 
In the limit of photometric 
coverage over a wide range of wavelength and with a large number of narrow-band filters, 
the resulting template fitting photometric redshifts can be determined almost unambiguously 
for each galaxy, i.e. with small statistical uncertainties, little ambiguity between different 
combinations of galaxy type and redshift, and almost full completeness. For less optimal 
photometric data, e.g.~when photometry is available only in a smaller range of bands, 
degeneracies between redshift and galaxy type for a given set of observed fluxes allow only 
for a probabilistic description of the redshift of each source, which crucially depends on 
assumed priors regarding the prevalence of galaxy types as a function of redshift and luminosity.
This is the case particularly for ongoing and upcoming four-to-six-band surveys like the 
Dark Energy Survey (DES, \citealt{2015arXiv150705603J}), the Kilo Degree Survey (KiDS, \citealt{2015MNRAS.454.3500K}), the Hyper-Suprime Cam Survey (HSC, \texttt{http://hsc.mtk.nao.ac.jp/ssp/}), and LSST \citep{2009arXiv0912.0201L}. 

An alternative approach, which we will refer to as \emph{empirical}, is to compare each source galaxy to a reference sample of galaxies with known (spectroscopic) redshift. The redshift $z$, or redshift 
distribution $p(z)$, assigned to the source galaxy is found from the redshifts of reference 
galaxies similar to it in terms of their photometric properties. Given a sufficiently large 
reference catalogue of galaxies selected in the same way as a photometric sample, the redshift 
distribution of the latter can be inferred from the redshift distribution of the former, 
both for individual sources and for the overall sample. 

In practical application, a number of effects can lead to sub-optimal or biased results in empirical methods, e.g. because
\begin{itemize}
\item photometric information on source galaxies is noisy and/or available only in few bands. When galaxies 
      of a range of redshifts have indistinguishable observed properties, selection of sources 
      and weighting by the amplitude of their lensing signal are not possible in a statistically 
      optimal way. For small reference catalogues or surveys, poor photometry also exacerbates cosmic variance.
\item selection effects in the source catalogue are not applied to the reference sample. 
      Because weak lensing measurements are often limited by noise due to the intrinsic shape 
      of background galaxies, one typically uses source samples that extend beyond the magnitude 
      limit where the catalogue is complete. In addition, shape measurement of galaxies has 
      success rates significantly below unity, even for galaxies with highly significant 
      photometric detections. Both of these selection effects depend on the size and other 
      morphological properties of source galaxies. If, at a given position in colour-magnitude 
      space, these properties correlate with redshift, biases are introduced.
\item selection effects are present in the reference catalogue that do not apply to the source 
      sample. Most spectroscopic surveys have selected targets by a combination of colours, 
      magnitude and sometimes additional morphological properties. Biases are introduced when 
      some of these do influence the redshift distribution but are not taken into account when 
      estimating $p(z)$, e.g. because they are not measured for the source galaxies. However, 
      even when the target selection is applied to the source galaxy sample, the incompleteness 
      of spectroscopic surveys and the fact that, especially at the faint end, success of 
      spectroscopic redshift recovery depends strongly on galaxy type and redshift (e.g.~on whether prominent emission lines fall inside the accessible wavelength range of a spectrograph), can cause additional biases.
\end{itemize}

The expectation value of $\beta$ for a lens redshift $z_d$ is estimated from $p(z)$ as
\begin{equation}
\langle\beta\rangle(z_d)=\int\frac{D_{ds}(z,z_d)}{D_s(z)}\;p(z)\;\rm{d}z \; .
\end{equation}
Biases in $p(z)$ therefore result in biases in $\langle\beta\rangle$.
For the purposes of lensing measurements of foreground structure, it is useful to describe systematic errors in 
an object's $p(z)$ by deviations of $\langle\beta\rangle$ from its true mean for a sample of 
source galaxies.

In this work, we quantify these systematic effects on empirical $p(z)$ estimation. 
To this end, we develop a simple colour-magnitude decision tree in \autoref{sec:method}, 
based on a magnitude limited reference sample with spectroscopically calibrated photometric 
redshifts from CFHTLS Deep optical and WIRDS near-infrared photometry. 
In \autoref{sec:bias}, we measure the effect of selection biases in spectroscopic subsamples, and due to the success of object detection and shape measurement in noisy data.  We conclude in \autoref{sec:conclusions}.
We describe three straightforward applications of the methodology presented in this paper to quantify the statistical power (\autoref{sec:variance}) and cosmic variance (\autoref{sec:cosmicvariance}) of empirical photometric redshifts as a function of bands used, and estimate biases in photometric redshifts from lensing magnification in \autoref{sec:magnification}.

All magnitudes used in this work are defined in the AB system and for CFHT Megacam / 
WIRCam filters $u^{\star}$~(u.MP9301), $g'$~(g.MP9401), $r'$~(r.MP9601), $i'$~(i.MP9701), 
$i'_2$~(i.MP9702, denoted $y$ elsewhere), $z'$~(z.MP9801), $J$~(J.WC8101), $H$~(H.WC8201) and $K_s$~(Ks.WC8302). 
Cosmological distances for the scaling of lensing signal amplitudes are calculated in a flat 
$\Lambda$ cold dark matter cosmology with $\Omega_m=0.27$.

\section{Method} \label{sec:method}

Photometric redshift methods aim to characterize the probability density function (PDF) of 
the redshift of a galaxy, $p(z)$, or more specifically the conditional PDF w.r.t.~its observed 
photometric properties. A useful frequentist understanding of the conditional $p(z)$ is that it is the 
distribution of true redshifts one would find in a complete spectroscopic follow-up of a 
large number of galaxies that look similar to the one in question in terms of its 
photometric properties.

Empirical redshift methods estimate a galaxy's $p(z)$ from the distribution of known true 
redshifts of similar galaxies. \emph{Similar} in this case refers to the observed properties 
of the galaxy in question, e.g.~magnitudes in the filter bands of some survey. Methods differ 
by how they establish similarity, e.g.~by interpolation of a (complicated) functional form 
such as for the case of artificial neural networks, kernel averaging in the space of 
observed parameters, or classification into subsamples. The optimal use of reference catalogues for redshift estimation has been explored with various statistical and machine learning techniques \citep[e.g.][]{1995AJ....110.2655C,2003MNRAS.339.1195F,2004PASP..116..345C,2008MNRAS.390..118L,2013MNRAS.432.1483C,2015arXiv150700490S,2015MNRAS.449.1275H,2015MNRAS.452.3710R,2016A&C....16...34H}. 

In the following sections, we describe the framework used in this work to test selection 
effects in empirical redshift estimation with a simple colour-magnitude decision tree applied 
to a magnitude-limited reference sample of galaxies with high-quality photometric redshift estimates. 

\subsection{Construction of colour-magnitude decision tree}
\label{sec:build}

As a framework for investigating systematic effects of empirical redshift methods, we develop 
a simple scheme based on classification of a magnitude-limited reference sample of galaxies 
using a decision tree in colour-magnitude space. By taking care to treat statistical limits 
of photometric measurements of colours correctly, our scheme yields unbiased estimates of $p(z)$ 
in the absence of selection effects. 

The purpose of the scheme is to describe the dependence of redshift on colour and magnitude adequately, yet with minimal complexity and in a way that allows systematic studies of errors in $p(z)$ estimation. 

To this end, for colour-magnitude information $(c_1, \ldots, c_n, m)$, we define a set of 
boxes (i.e., hyper-rectangles) $B_i$ in $(n+1)$-dimensional space as 
\begin{equation}
B_i=]c_{1,i}^{\rm min},c_{1,i}^{\rm max}]\times\ldots\times]c_{n,i}^{\rm min},c_{n,i}^{\rm max}]\times]m_i^{\rm min},m_i^{\rm max}] \; .
\end{equation}
Each source that falls into $B_i$ is assigned the observed distribution of redshifts of 
reference galaxies in $B_i$ as its $p(z)$.

To build the decision tree from a reference catalogue of galaxies with known redshift (and, therefore, $\beta$ 
for a given lens redshift $z_d$), we
\begin{enumerate}
\item split the sample into two equally sized subsamples 1 and 2 at the median of each of the $n+1$ properties, then
\item select the property where the split yields the maximal value of $\langle\beta\rangle_1^2+\langle\beta\rangle_2^2$, and finally
\item accept the split unless a stopping criterion is fulfilled and continue to both subsamples further.
\end{enumerate}
The condition (ii) for the optimal split can be understood from considering the recovered 
lensing signal-to-noise ratio (S/N) from the optimally weighted source sample (cf.~\autoref{sec:variance}). 

\begin{figure}
  \includegraphics[width=\linewidth]{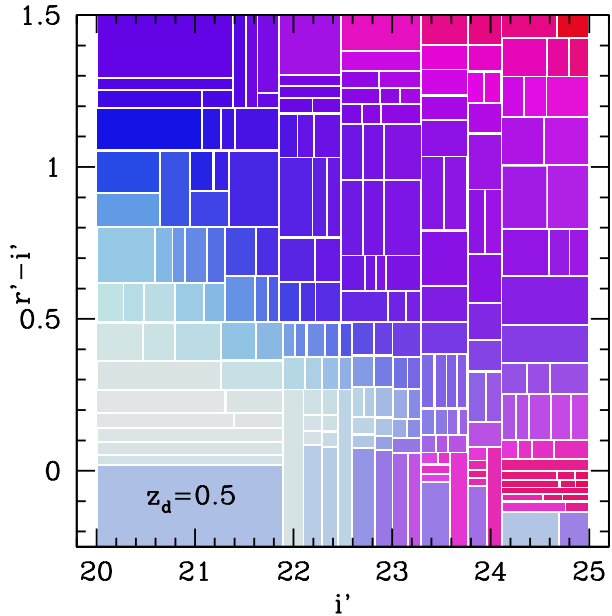}  
  \caption{Decision tree for estimation of $\beta$ at $z_d=0.5$ from $i',r'-i'$ colour-magnitude
  information of source galaxies with $20<i'<25$. The colour of each box indicates the mean 
  value of $\beta$ (from 0, cyan, to the maximum, red) and the saturation the level of recovery of lensing S/N relative to selection and weighting by spectroscopic redshifts (cf. \autoref{sec:variance}).}
\label{fig:tree}
\end{figure}

\red{A split is performed only if all of the following criteria are met:}
\begin{itemize}
\item the split is such that for \red{all} of the galaxies in the reference subsample, \red{all} of the 
      relevant bands \red{defining the new subsamples are} measured with a S/N of flux of \red{at least} 10,
\item the difference between $\langle\beta\rangle_1$ and $\langle\beta\rangle_2$ is \red{larger} 
      than twice its uncertainty estimated from the sample variance \red{in the subsample, and}
\item the subsamples contain \red{at least} 100 reference galaxies.
\end{itemize}
Once, for a subsample, splits at the median of \red{none} of the properties meet these criteria, we stop and use its limits as a colour-magnitude box. \autoref{fig:tree} illustrates the tree structure in 2D generated from $i',r'-i'$ information for a lens redshift of $z_d=0.5$.

We note that, while simplistic, the colour-magnitude decision tree employed in this work 
yields a self-consistent $p(z)$ estimate: applied to the reference sample itself (or a 
representative subsample thereof), the estimated $p(z)$ or $\beta$ has the exact (expectation) 
value. And given a magnitude-limited catalogue of a certain depth and set of observed bands, 
it allows us to build appropriate decision trees for any shallower magnitude-limited sample 
observed in a subset of the bands. This is a potential advantage over simple kernel density estimation (where 
non-linear dependence of mean redshift on photometry over the range of the kernel causes a bias) 
or machine learning schemes. While it is conceivable that more elaborate methods yield smaller 
statistical uncertainties, i.e.~in terms of the width of the $p(z)$ of individual galaxies, the decision tree 
method is a useful baseline for an empirical method that is unbiased when applied 
to reference catalogues and source samples in the absence of selection effects.

One limitation of the method is the effect of noise in the photometry of source galaxies for which the $p(z)$ should be estimated. With noisy flux measurements, measured and true magnitudes of galaxies can correspond to different colour-magnitude boxes. \red{This is fully accounted for when galaxies in the reference catalogue and galaxies in the sample to which the method is applied are scattered between their true colour-magnitude box and surrounding boxes with the same probability. One way of achieving this would be to re-shuffle} the reference galaxies between boxes according to noise added to their flux measurements to match the higher flux errors of the respective source galaxy. For all tests run in this work, however, we use the reference sample itself as source galaxies with the exact same flux measurements for building the tree, estimating the $p(z)$ of each cell, and assigning sources to cells, and such a procedure is therefore not necessary.

\red{We note that estimating the $p(z)$ of a (sub)sample of the galaxies used to build and populate the tree is advantageous for estimating the impact of selection effects, as in \autoref{sec:specbias}, \ref{sec:photbias}, and \ref{sec:shapebias}, or shifts in photometry, as in \autoref{sec:magnification}, since it nulls the effect of sample variance. For evaluating the quality of the tree method for photometric redshift estimation, it would be preferable to use separate (\emph{training} and \emph{validation}) samples for populating the tree and finding the redshift distributions of. Given the large number of reference galaxies, we expect that our use of the full sample for both has only a minor impact on the results of \autoref{sec:variance} and \autoref{sec:cosmicvariance}.}

\subsection{Reference catalogue}
\label{sec:refcat}
The empirical method developed and tests performed in this work rely on a reference sample 
of galaxies with known (true) redshifts and colour-magnitude information in a superset of the 
bands available for the galaxies we would like to estimate the $p(z)$ for. The reference 
catalogue needs to select galaxies only by their (extinction-corrected) total magnitude in a 
single band, i.e. should make no selection based on colour, size, type or observing conditions. 
Since no large and deep enough sample with spectroscopic redshifts is available that matches 
these criteria, we resort to a sample with high-quality photometric redshifts. The underlying
assumption of the tests to follow is that these are the \emph{true} redshifts of the reference
sample of galaxies. The tests are defined, however, to be almost independent in their result
from the validity of this assumption.

For the purpose of getting a useful sample of galaxies, we use imaging data in $u^{\star}g'r'i'i'_2z'$ from the 
Canada-France-Hawaii Telescope Legacy Survey (CFHTLS) Deep fields,\footnote{\texttt{http://www.cfht.hawaii.edu/Science/CFHLS/ cfhtlsdeepwidefields.html}} 
overlapping with $JHK_s$ data from the WIRCam Deep Survey (WIRDS, \citealt{2012A&A...545A..23B}). 

The catalogue creation follows the procedure described in \citet{fabrice}. 
We use the \textsc{SExtractor} software \citep{1996A&AS..117..393B} in dual-image
mode to extract the photometric fluxes, magnitudes and errors. In order to
obtain meaningful colour estimates, we at first adjust the PSF in the
different observed filters by degradation to the value of the worst band
(which is in general $u^{*}$). We do so by convolving the co-added images with an
appropriate global Gaussian kernel so that the measured stellar colours in
average no longer depend on the considered aperture. We then use the
unconvolved $i'$-band as the detection band and extract the photometric
fluxes from the convolved images, making use of weight images. We
extract all objects that are at least 2$\sigma$ above the background on at
least four contiguous pixels.
Unfortunately, the image zeropoints provided with CFHTLS Deep are of limited accuracy,
which leads to field-to-field variations of colour. We account for
this by applying a stellar locus regression, making use of the predicted
stellar colours of the Pickles star library (1998), thus homogenizing the
colour estimates in the different pointings.
In the next step we obtain preliminary photometric redshift estimates
using the template-fitting Photo-Z code of \citet{2001defi.conf...96B}, which has
been successfully applied in a variety of lensing-related contexts (e.g. \citealt{fabrice}, \citealt{2012MNRAS.420.1384S}).
In the final steps we then verify the achieved photometric redshift accuracy
by comparison with the available spectroscopic overlap (see section 2.2.1)
and use the information to apply the final spectroscopic calibration to our
photometric redshift sample, obtaining a very good photometric redshift
accuracy of $\sigma_{\Delta z/(1+_{z_{\rm spec}})} = 0.026$ and $\Delta z/(1+z) = 0.031$ in scatter and outlier rate of $\eta = 2.1$ per cent for galaxies with $i' \le 24.0$ (cf. \autoref{fig:specz}).

Combining all CFHTLS Deep fields, this yields over one million galaxies detected down to $i\approx27.5$. In order to ensure
completeness, we however apply the following cuts.

We exclude objects above a level of background noise in any of the bands 
such that our sample has a signal-to-background-noise ratio of at least 10 in 
apertures of 8~pix (1.5~arcsec) diameter for $u^{\star},g',r'<26.5$, $i'<26$, $i'_2,z'<25.5$. 
Infrared coverage is less uniform, such that while we 
require that objects are not masked in the WIRDS data, we include all area above limiting magnitudes of $J<20$, 
$H<19.5$ and $K_s<19$. We ensure that for objects fainter than these magnitude limits, only an 
upper limit in flux at this magnitude is used for $p(z)$ estimation. We do \emph{not}, however, 
reject objects with fluxes below these limits (i.e. drop-outs).

\begin{figure}
  \includegraphics[width=\linewidth]{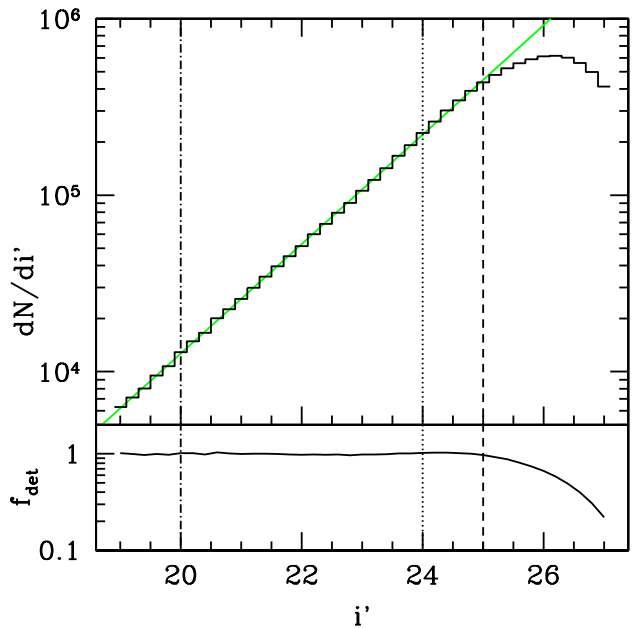}  
  \caption{Upper panel: Distribution of apparent \emph{i'} band magnitude of galaxies detected in the CFHTLS Deep fields. Green line indicates a power-law fit in the range $19<i'<22$. Lower panel: Completeness of object detection estimated relative to the power-law fit. Vertical lines indicate cuts for deep ($20<i'<25$) and shallow samples ($20<i'<24$) used later.}
\label{fig:lf}
\end{figure}

\begin{figure}
  \includegraphics[width=\linewidth]{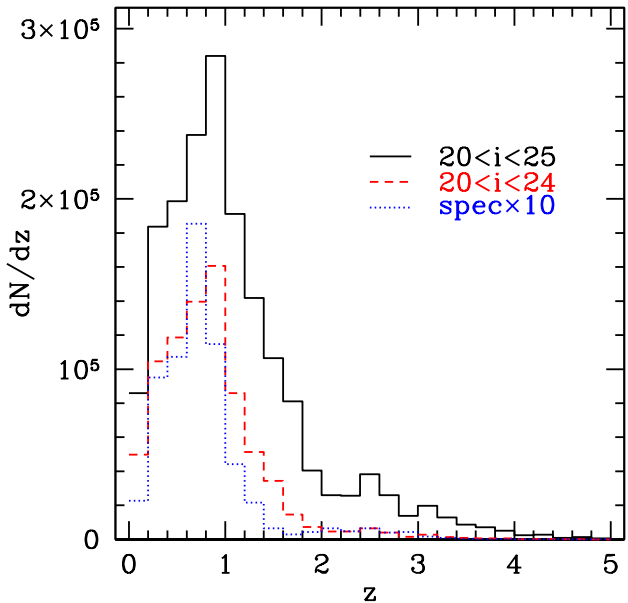}  
  \caption{Distributions of template-fitting redshift point estimate for deep $20<i'<25$ sample (black, solid line), bright $20<i'<24$ sample (red, dashed line) and 
  spectroscopic subsample (blue, dotted line, multiplied by 10 to match scale).}
\label{fig:pz}
\end{figure}

\begin{figure}
  \includegraphics[width=\linewidth]{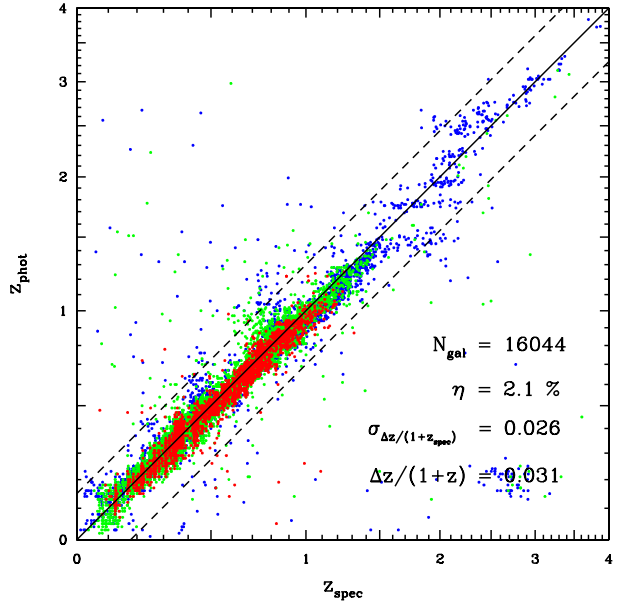}  
  \caption{Photometric redshift point estimates of spectroscopic galaxies in CFHTLS Deep compared to their spectroscopic redshifts. The plot shows all galaxies from the VVDS-Deep, VIPERS, zCOSMOS-bright, zCOSMOS-deep, VUDS and DEEP2 spectroscopic surveys with valid CFHTLS Deep and WIRDS photometry in all bands and $i'<24$. \red{Red, green and blue symbols indicate that the best-fitting templates correspond to early-type, late-type and starburst galaxies in the classification of \citet{2005ApJ...631..126D}}.}
\label{fig:specz}
\end{figure}

The only cut dependent on galaxy properties is that we exclude objects brighter 
than $i'=20$, where $\mathrm{d}N/\mathrm{d}i'<5000$, or fainter than $i'=25$, where 
$\mathrm{d}N/\mathrm{d}i$ begins to deviate from a power-law (cf.~\autoref{fig:lf}).
A small number of objects (54 in total) need to be excluded from the sample because 
the template fit fails. This combination of cuts yields 348,601 galaxies in the four CFHT 
pointings (cf.~\autoref{tab:samples}). The redshift distribution of the deep and bright sample are shown in Figure \autoref{fig:pz}.

We note that the deep sample of $i'<25$ is complete over the magnitude range of shape catalogues 
used for the CFHT Lensing Survey \citep[CFHTLenS,][]{2012MNRAS.427..146H,2013MNRAS.429.2858M} as well as the 
Weighing the Giants \citep[WtG,][]{2012arXiv1208.0597V,2012arXiv1208.0605A} and COnstrain Dark Energy with X-ray clusters \citep[CODEX,][]{nathalia} dedicated cluster weak lensing surveys.
We define a bright subsample as $20<i'<24$,
more representative of shape catalogues in ongoing large-area surveys such as 
DES \citep{2015arXiv150705603J} or KiDS \citep{2015MNRAS.454.3500K}, consisting of 159,065 galaxies.

\red{Other publicly available, deep, high-quality photometric redshift catalogues, such as those of the COSMOS field \citep{2006A&A...457..841I,2009ApJ...690.1236I,2016arXiv160402350L}, the ESO Deep Public Survey \citep[ESO-DPS,][]{2005AN....326..432E,2006AuA...452.1121H}, or the ALHAMBRA survey \citep{2013arXiv1306.4968M} are less useful for the tests performed here due to a combination of smaller area or depth or lacking CFHTLS overlap. Particularly the overlapping COSMOS photo-$z$ catalogue of \citet{2016arXiv160402350L} is an excellent complement for tests of the influence of photo-$z$ uncertainties \citep[see][their Figure~3]{nathalia}, which are however not of great relevance to the tests of selection effects in this work.} 

\subsubsection{Spectroscopic samples}

A number of spectroscopic samples cover the CFHTLS Deep fields, and we list their details
in this section.

The VIMOS VLT Deep Survey (VVDS-Deep, \citealt{2005A&A...439..845L}), with one of its two 
fields overlapping D1, \red{targets a magnitude-limited sample at $I<24$. Its overall high spectroscopic completeness (78 per cent with high confidence 
redshifts) is a strong function of apparent magnitude, and decreases especially for $I>23$, with known colour dependence of the incompleteness \citep[cf.][their Fig.~7]{2015arXiv150705909B}.}

The VIMOS Public Extragalactic Survey (VIPERS, \citealt{2014A&A...562A..23G,2014A&A...566A.108G})
is covered by the CFHTLS Wide fields W4 and W1, of which D1 is a pointing. To increase the 
coverage of galaxies at high redshift, VIPERS uses a $u^{\star}g'r'i'$ colour cut in addition 
to a $i'<22.5$ magnitude limit. 

zCOSMOS \citep{2007ApJS..172...70L} partly overlaps D2 and consists of two separate sub-samples: 
one (zCOSMOS-bright) magnitude-limited at $I<22.5$ with $\approx90$ per cent success rate and a 
second sample (zCOSMOS-deep) of fainter galaxies pre-selected by a colour and $B$ band magnitude 
limit to be within the redshift range of $1.4<z<3.0$.

There are spectra from one pointing of the VIMOS Ultra Deep Survey (VUDS, \citealt{2015A&A...576A..79L,2016arXiv160201842T})
in the D2 field. These include very faint objects out to $i'>25$, which are, however, selected
from a complex scheme of colour and photo-$z$ cuts to optimize the yield of high redshift galaxies. 

The Deep Extragalactic Evolutionary Probe-2 (DEEP2) survey \citep{2013ApJS..208....5N} overlaps in part with D3 and, in addition to 
a magnitude limit at $R<24.1$, has used cuts in $BRI$ colours to pre-select galaxies with $z>0.7$.

From all these samples, we use only objects with high-confidence spectroscopic redshifts 
(quality flag 3 or 4 in the respective catalogue) and include secondary targets that happen 
to fall inside slit masks.

In summary, the only magnitude limited samples (although in a filter that differs slightly 
from $i'$) overlapping CFHTLS Deep are VVDS-Deep \red{down to $I\approx23$} and zCOSMOS at $I<22.5$. DEEP2 
and zCOSMOS-deep apply cuts that preferentially reject low-redshift sources, based on $B$ 
band that is not available from CFHTLS Deep photometry. The expectation is that this cut 
might bias the spectroscopic sub-sample to higher redshift than a colour-magnitude limited 
sample based on redder bands only.

\begin{table}
\centering
\caption{Photometric and spectroscopic samples in the CFHTLS Deep fields after the cuts described in \autoref{sec:refcat}.}
\label{tab:samples}
\begin{tabular}{lcc}
\hline
field & phot./spec. & \# \\
\hline
D1 & phot. & 96748 \\
VVDS-Deep & spec. & 3187 \\
VIPERS & spec. & 1728 \\
\hline
D2 (COSMOS) & phot. & 130759 \\
zCOSMOS-bright & spec. & 3022 \\
zCOSMOS-deep & spec. & 904 \\
VUDS & spec. & 109 \\
\hline
D3 & phot. & 64003 \\
DEEP2 & spec. & 3830 \\
\hline
D4 & phot. & 57091 \\
\hline
\end{tabular}
\end{table}

\section{Selection effects} \label{sec:bias}

An idealized empirical method of redshift estimation should return the redshift distribution $p(z|m,c,S)$
of a source estimated under the condition of its observed properties (e.g. colour-magnitude position $m,c$) 
and $S$ which encodes selection by additional properties, e.g. position in the sky or other features that 
influence the success of inclusion in the survey.

For a reference catalogue spanning the colour-magnitude region of interest but observed with 
a selection function $S_{\rm ref}$, an empirical algorithm can only be expected to return 
$p(z)=p(z|m,c,S_{\rm ref})$ (in the absence of re-weighting). This need not be an issue: for a magnitude-limited reference 
catalogue and a shallower, magnitude-limited survey, which differ by their footprints in the sky, 
$S\neq S_{\rm ref}$ (because of differences in depth and position), yet 
$p(z|m,c,S_{\rm ref})=p(z|m,c,S)$.\footnote{This is only strictly true under the assumption of noiseless $m,c$ and up to cosmic variance.}

Potentially more harmful is a selection function of the reference catalogue $S_{\rm ref}$ 
that contains colour-magnitude information not contained in $(m,c)$ (e.g. selection by a 
band unobserved in the survey or by the success of identifying features in the spectra that
allow a certain redshift determination). Likewise, when the reference catalogue is truly magnitude
limited but the shallower survey contains only a fraction of objects near the detection limit for which
photometric measurements were successful, it is imaginable that 
$p(z|m,c,S_{\rm ref})\neq p(z|m,c,S)$.

In the following sections, we test for selection effects that significantly change the mean value of $\beta$ and the
redshift distributions of galaxy samples at a fixed $(m,c)$. 

For the latter, in each cell of the colour-magnitude decision tree we compare the distributions of eight-band template-fitting photo-$z$ of two samples, where one has an additional selection effect applied. For testing
the selected subsamples for compatibility of their $p(z)$ with the full sample, we run a Cram\'{e}r-von Mises (CvM) test. When ordering the galaxies in a colour-magnitude box by their photometric redshift point estimate,
galaxies with spectroscopic confirmation should be interspersed randomly throughout the list. The CvM statistic $T$ quantifies to which degree this is the case.

For two sets $u$ and $v$ of $N$ and $M$ values (in our case, of photometric redshift) with ranks 
$u_i, i=1,\ldots,N$ and $v_i, i=1,\ldots,M$ 
in the ordered, combined set, the CvM statistic $T$ is defined as
\begin{equation}
T=\frac{N\sum_{i=1}^N(\hat{u}_i-u_i)^2+M\sum_{i=1}^M(\hat{v}_i-v_i)^2}{NM(N+M)}-\frac{4NM-1}{6(N+M)} \; .
\label{eqn:cvm}
\end{equation}
In the case that two or more galaxies in our sample share equal redshift (which is common due to
the gridded photometric redshift estimates and the fact that, in our tests, one sample is always a subsample of the other), 
we assign the mean of their combined ranks to all equal redshift objects.

The $T$ statistic for a cell in colour-magnitude space can be interpreted in terms of 
its $p$-value, i.e. the probability of finding a value of $T$ higher than what has been found
given that $u$ and $v$ are random subsamples from the same population of galaxies. 
Since the analytical calculation of the $p$-value is computationally expensive for large sample
sizes \citep{JSSv017i08}, we use Monte-Carlo realizations of random subsamples to estimate $p$.
\red{This ensures that the $p$ values are uniformly distributed between 0 and 1 for random subsamples of the same size.}

\subsection{Spectroscopic selection}

\label{sec:specbias}

\begin{figure*}
\subfigure[all spectroscopic data]{
  \includegraphics[width=0.48\linewidth]{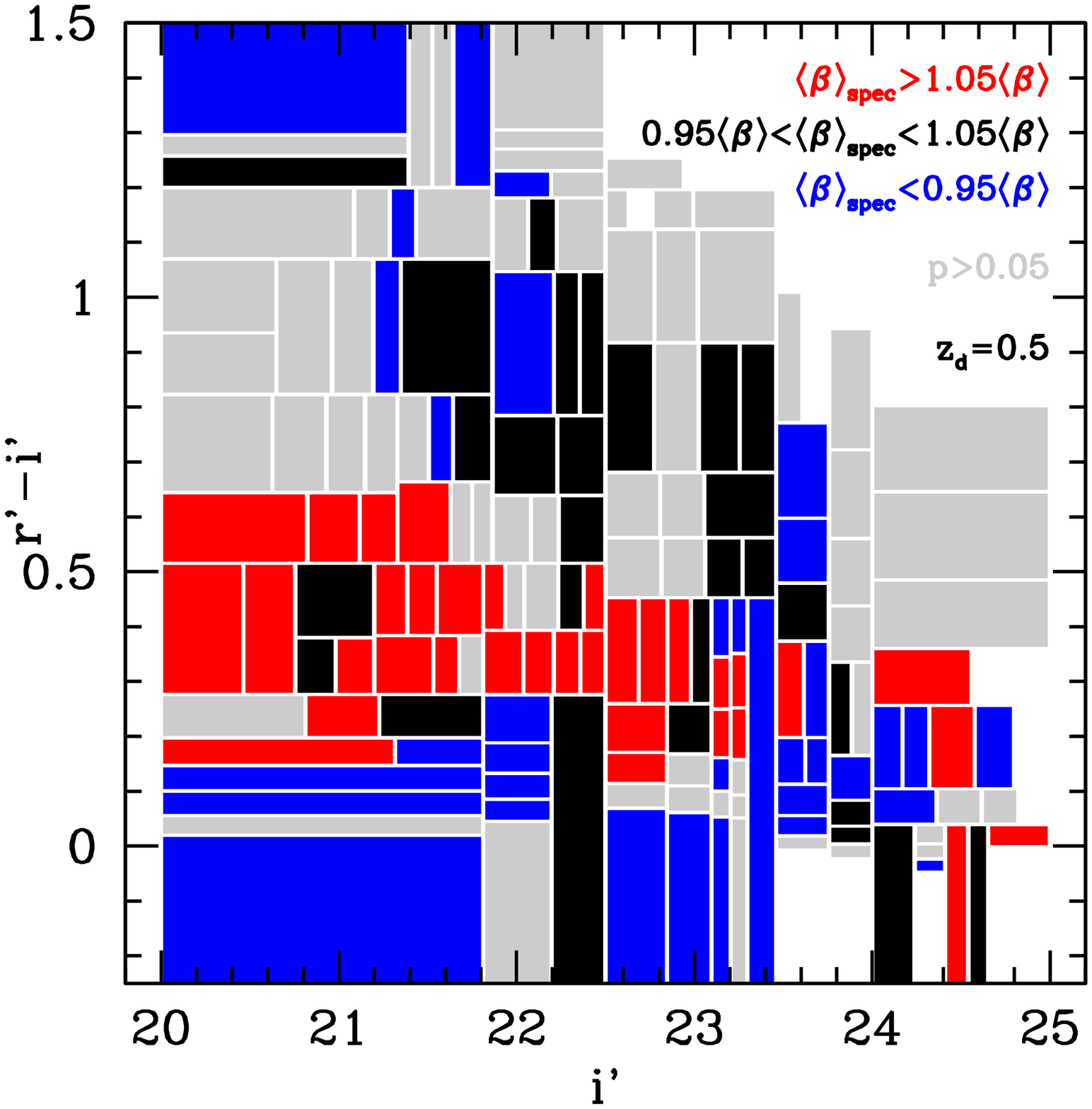} 
}  
\subfigure[magnitude limited spectroscopic samples only]{
  \includegraphics[width=0.48\linewidth]{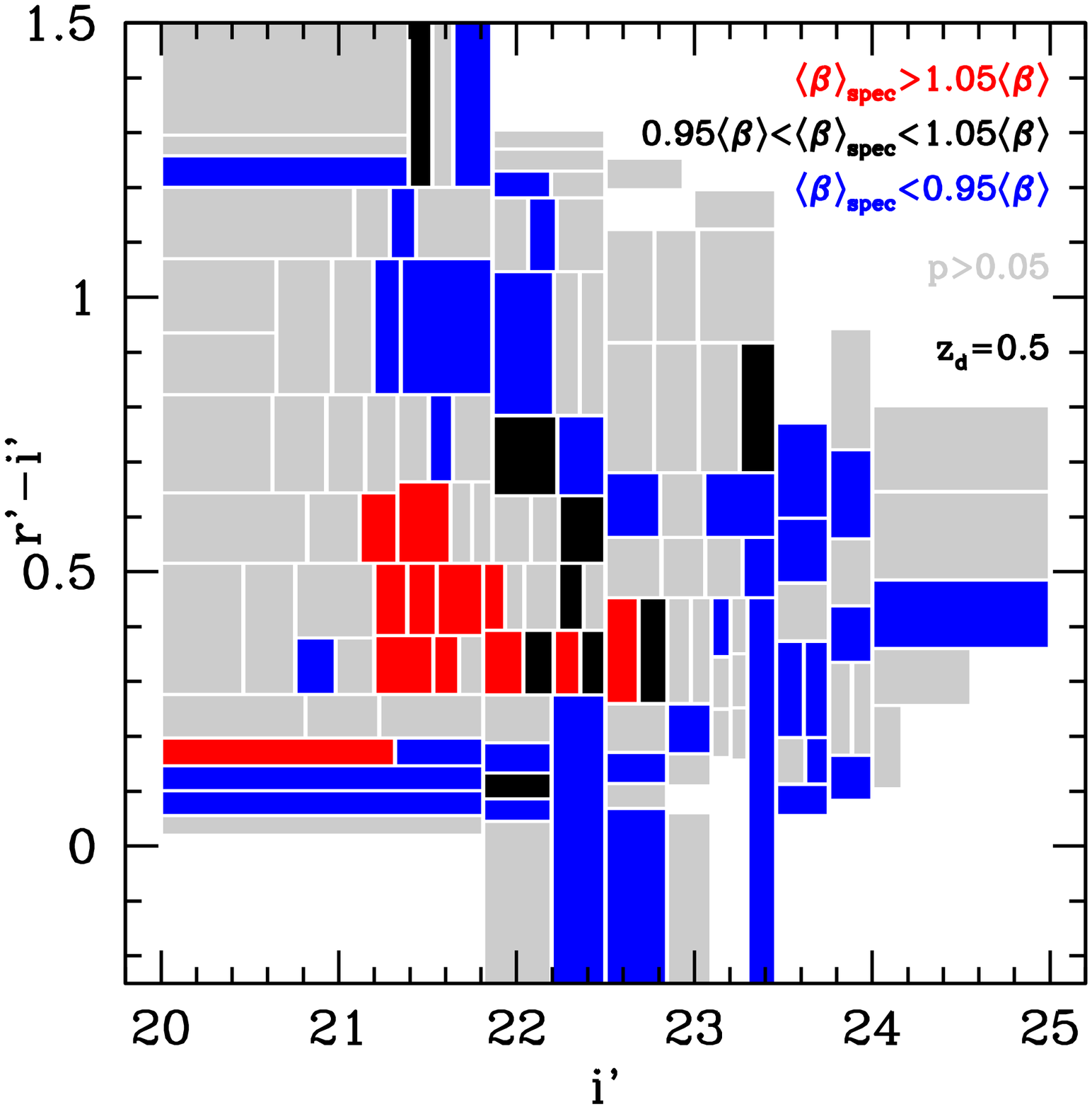} 
}
  \caption{Cram\'{e}r-von Mises test of $p(z_{\rm phot})$ of all objects in each 
  colour-magnitude cell vs. the selection of objects for which there is spectroscopic follow-up. Included are cells in $i',r'-i'$ space with at least 10 spectroscopic objects. $p$-values are calculated for each of the spectroscopic surveys individually on the respective CFHT pointing alone and combined using \autoref{eqn:pcombine} to avoid cosmic variance. Cells with significant differences in the distributions ($p<0.05$) are coloured according to the difference in mean $\beta$ between photometric and spectroscopic objects according to their photometric redshift. \textit{Left-hand panel:} full spectroscopic sample. Cuts at $i'=22.5$ and $24$ have been imposed. \textit{Right-hand panel:} magnitude-limited spectroscopic samples only.}
  \label{fig:cvm}
\end{figure*}

Selection biases in spectroscopic samples can exist either because the target selection is based on 
colours not measured in the subset of bands used for the empirical $p(z)$ estimate or because
the success of spectroscopic redshift determination depends on properties not fully determined by the 
photometric information. While these are particularly evident in some cases, e.g. the PRIMUS spectra with a redshift-based selection \citep{2013ApJ...767..118C} that could not be utilized for the purpose of DES redshift calibration \citep{2015arXiv150705909B}, less obvious spectroscopic selection effects commonly go unchecked.

To test this as robustly as possible, we apply the CvM statistic of \autoref{eqn:cvm}. We use
the same eight-band photo-$z$ estimates of redshift of both samples when comparing the distributions
to ensure that disagreement is not due to imperfections in the photo-$z$. In order to 
exclude cosmic variance or differences in magnitude calibration as a source of false positives 
for the disagreement, we run the comparisons individually for each of the spectroscopic samples, 
using in each case only photometric galaxies from the same CFHTLS Deep pointing.  

Given $k$ independent tests (made with independent spectroscopic samples for the same 
colour-magnitude cell) of the same null hypothesis (that the distributions of photo-$z$ of the full sample
and spectroscopic subsample agree) that yield $p$-values $p_i$, $i=1,\ldots,k$, 
Fisher's method allows us to combine these into a random variable
\begin{equation}
X=-2\sum_{i=1}^k\ln p_i \;.
\label{eqn:pcombine}
\end{equation}
$X$ is $\chi^2$-distributed with $2k$ degrees of freedom, and the combined $p$-value of the
null hypothesis is $p(\chi^2_{2k}>X)$.

\autoref{fig:cvm} shows regions in $i',r'-i'$ space for which the null hypothesis is rejected
($p<0.05$), i.e.~where spectroscopic subsamples have different photometric properties in the
unused $u^{\star}g'z'JHK_s$ bands that lead to an incompatible photo-$z$ distribution. The 
disagreement is widespread for the full sample. Limiting the spectroscopic sample to the
magnitude limited VVDS-Deep and zCOSMOS-bright samples only, a smaller range of colour-magnitude
space shows significant deviations of redshift distributions between the two samples, 
although there is still clear evidence for selection effects in some regimes. These are likely due to incomplete success of these surveys at faint magnitudes.

While the case of $i',r'-i'$ is useful for illustration, realistic lensing surveys commonly use more than two bands for defining background galaxy samples and estimating redshift distributions. Selection biases on $p(z|c,m,S)$ are reduced in this case, since some of the selection is absorbed from $S$ into $c$.

\begin{figure*}
\subfigure[]{
  \includegraphics[width=0.48\linewidth]{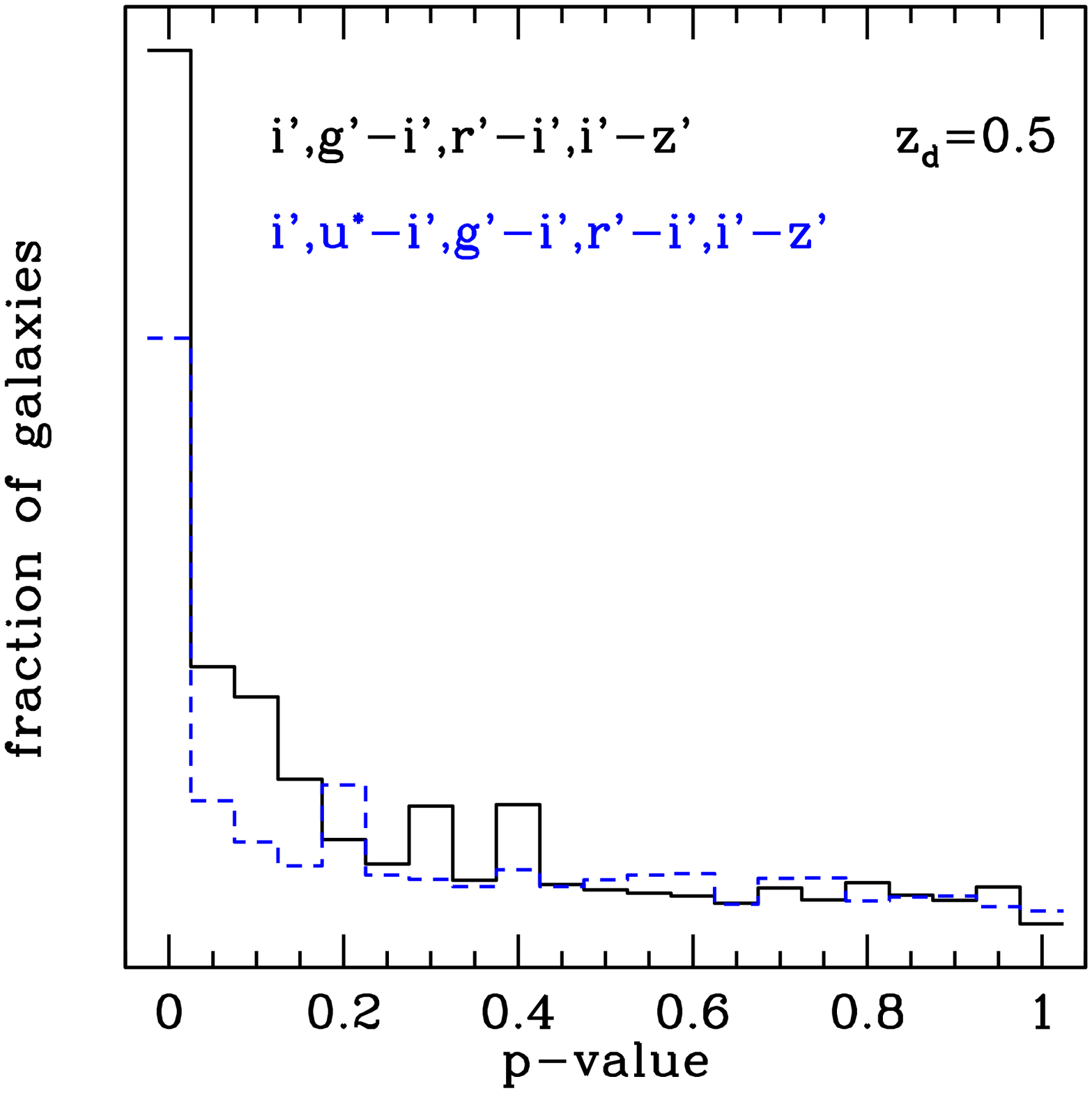}
    \label{fig:phist}
  }
  \subfigure[]{
    \includegraphics[width=0.48\linewidth]{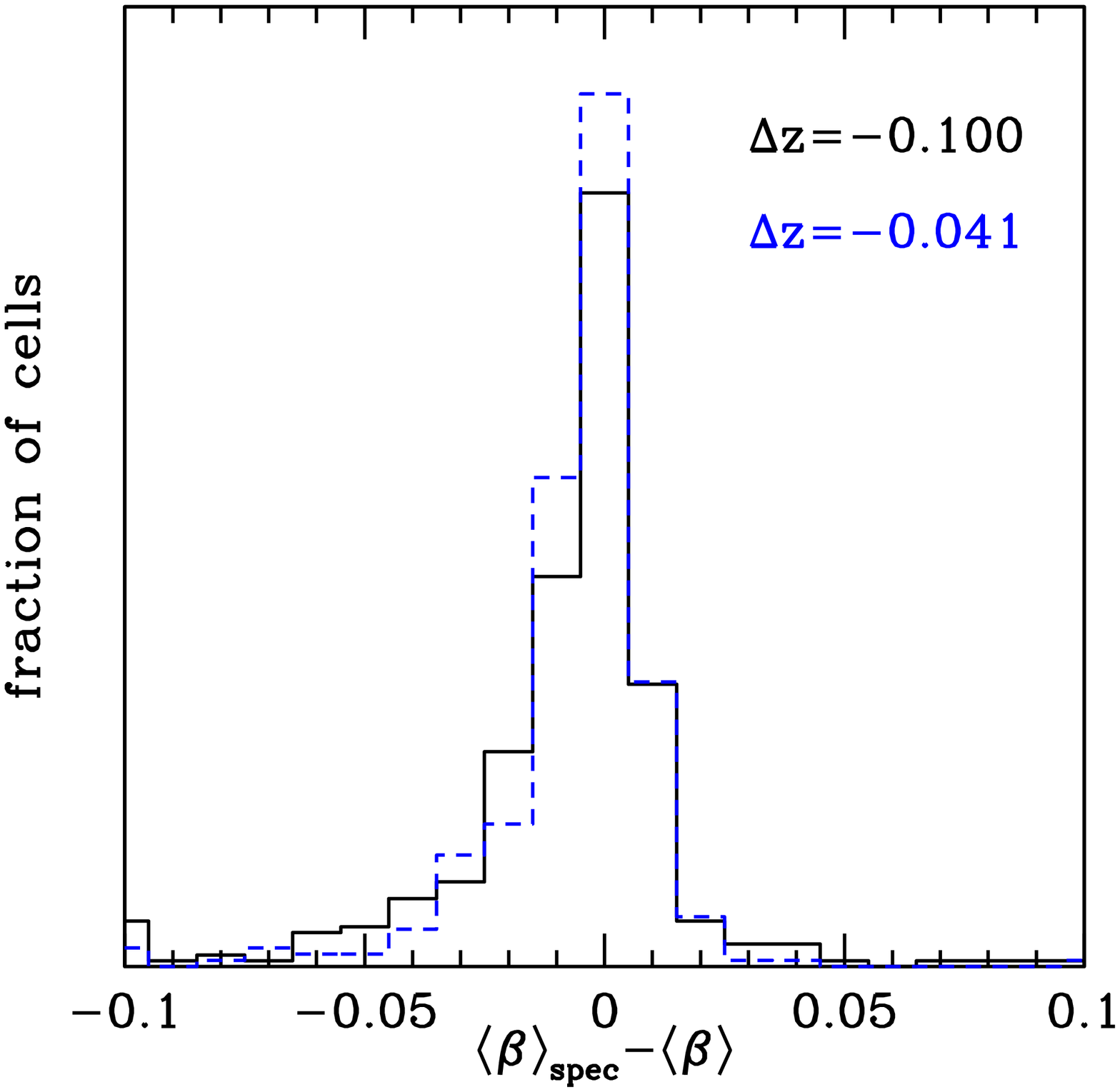} 
      \label{fig:dbhist}
  } 
  \caption{\textit{Left-hand panel:} Histogram of $p$-values of Cram\'{e}r-von Mises test comparing the $p(z_{\rm phot})$ of all objects in each colour-magnitude cell to the selection of objects for which there is spectroscopic follow-up. From the fraction of galaxies lying in cells where $p>0.5$, we estimate that 30 per cent (43 per cent) of galaxies are represented without selection bias in the spectroscopic catalogues when splitting them by $g'r'i'z'$ ($u^{\star}g'r'i'z'$). Only colour-magnitude cells with at least 10 spectroscopic objects are used. \textit{Right-hand panel:} Histogram of difference in mean $\beta$ estimated from photometric redshifts of spectroscopic vs. all objects in a colour-magnitude cell.}
\end{figure*}

Figure~\autoref{fig:phist} shows the distribution of $p$-values of the CvM test in $g'r'i'z'$ and $u^{\star}g'r'i'z'$ colour-magnitude space (as two common example of the many possible combinations of bands), weighted by the number of galaxies in the respective cell from a magnitude-complete sample at $20<i'<25$. While the addition of bands ($u^{\star}$ band in particular) reduces selection effects considerably, a large fraction of galaxies still reside in regions of colour-magnitude space where spectroscopic subsamples are significantly non-representative.

The distribution of residuals in $\langle\beta\rangle$ between full sample and spectroscopic subsample in Figure~\autoref{fig:dbhist} reveals an asymmetric tail towards lower redshift of the spectroscopic subsample.

\begin{figure}
  \includegraphics[width=\linewidth]{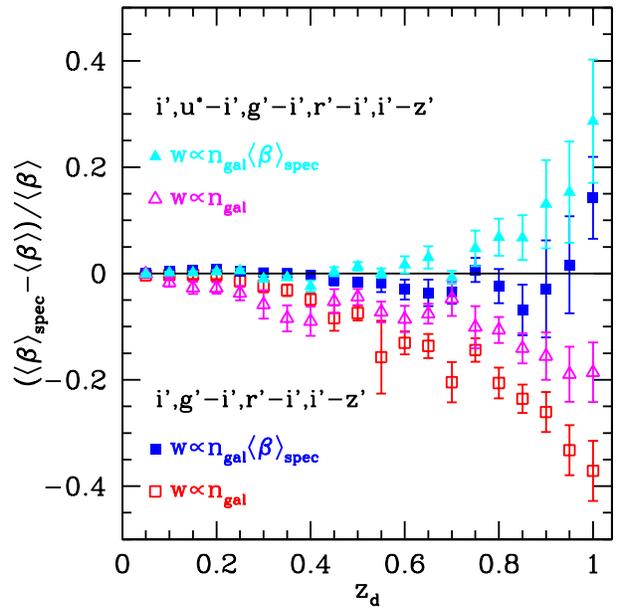} 
  \caption{Relative bias in mean $\beta$ (i.e. in estimated $\Delta\Sigma$) between spectroscopic subsample and magnitude-limited source sample at $20<i'<25$. Blue (cyan) filled square (triangle) symbols and red (magenta) open symbols show minimum variance lensing weighting and uniform weighting of colour-magnitude cells in $g'r'i'z'$ ($u^{\star}g'r'i'z'$), respectively. Cells without spectroscopic galaxies are excluded from the source sample.}
  \label{fig:betadiff}
\end{figure}

The overall impact of spectroscopic selection is best described by the (relative) bias on the mean value of $\beta$ of the spectroscopic subsample relative to the full sample, which would directly correspond to biases on $\Delta\Sigma$ in a lensing analysis that calibrates its source $p(z)$ from spectra. \autoref{fig:betadiff} shows this as a function of lens redshift and for two common weighting schemes of source galaxies: one where each source has the same weight in a mean shear analysis \citep[e.g. as in a non-tomographic cosmic shear analysis or in][]{2016arXiv160407871K} and one where sources are weighted by their estimated $\langle\beta\rangle$ for a minimum-variance measurement of $\Delta\Sigma$ \citep[e.g.][]{Sheldon04.1}.

The main effect is an underestimation of $\beta$, especially for intermediate to large lens redshifts $z_d$. Minimum-variance weighting by $\langle\beta\rangle_{\rm spec}$ alleviates (in fact, sometimes overcompensates) the bias by giving less weight (more weight) to parts of colour-magnitude space in which the spectroscopic sample underestimates (overestimates) the mean value of $\beta$. In both weightings, however, the bias in $\beta$ is of problematic amplitude, especially for $z_d>0.4$. The amplitude of the effect gets somewhat smaller with the addition of $u^{\star}$ band, although the comparison rests on different subsamples due to the exclusion of colour-magnitude cells without spectroscopic galaxies.

We calculate an alternative metric, the mean redshift bias $\Delta z=\langle z\rangle_{\rm spec}-\langle z\rangle_{\rm all}$ of all galaxies in the magnitude-limited sample from cells with at least one spectroscopic object. For $u^{\star}g'r'i'z'$ ($g'r'i'z'$) and the colour-magnitude decision tree generated for $z_d=0.5$, we find $\Delta z=0.839-0.880=-0.041$ ($-0.100$, respectively), at levels exceeding the requirements of ongoing lensing surveys \citep[e.g.][]{2016arXiv160707910S}.
 
\red{Importantly, these selection biases cannot be corrected by any weighting scheme that is based on the observed magnitudes in a $(u^{\star})g'r'i'z'$ survey: even at the same observed colour-magnitude (i.e.~in one colour-magnitude cell), the redshift distributions of spectroscopic and full sample differ.}

\red{We note that these results are found with the underlying assumption that the eight-band photometric redshifts from the CFHTLS and WIRDS photometry are the \emph{true} redshifts of the galaxies in our reference sample. Spectroscopic and magnitude-limited samples differ in their eight-band template-fitting $p(z)$, e.g.~in a $g'r'i'z'$ colour-magnitude cell. This means that in that cell, the distribution of the remaining $u^{\star}JHK_s$ magnitudes of the two samples differ in a way that impacts the template fitting. This is a conservative definition for the purpose of detecting selection effects: additional spectroscopic selection effects that are not recognizable with this set of 8 filters or the templates used are possible. Our comparison of the CFHTLS eight-band with spectroscopy and COSMOS 36-band photometric redshifts \citep{2016arXiv160402350L} in \citet{nathalia} indicates, however, that in most of colour-magnitude space the eight-band information is sufficient for highly accurate redshift estimation.}

\subsection{Photometric incompleteness}
\label{sec:photbias}

Faint sources near the detection limit dominate the background galaxy samples in contemporary imaging surveys, in terms of their number and their contribution to lensing signal-to-noise. To reach sufficient background galaxy number densities for weak lensing studies, it is typically necessary to include sources at magnitudes where the completeness is below unity. As an example, \citet{2012arXiv1208.0605A} apply a relatively conservative cut at the magnitude $m_t$ where $dN/dm$ turns over. In the photometric reference catalogue we use, however, for a cut at magnitudes $\approx m_t-1$ or fainter, the sample is not truly magnitude limited. 

\begin{figure}
  \includegraphics[width=\linewidth]{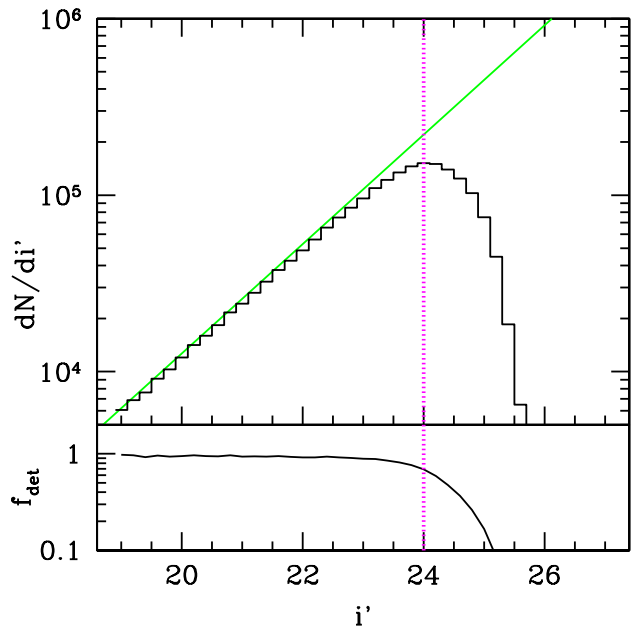}  
  \caption{Upper panel: Distribution of apparent \emph{i'} band magnitude of galaxies detected in the CFHTLS Deep fields after adding Gaussian noise as described in the text. Green line indicates a power-law fit to the full photometric sample in the range $19<i'<22$.  Lower panel: Completeness of object detection is estimated relative to the power-law fit. Vertical line indicates the peak of the histogram at $i'_t=24$. For comparison the figure is shown with the same axis scaling as \autoref{fig:lf}. }
\label{fig:lf24}
\end{figure}

\begin{figure}
  \includegraphics[width=\linewidth]{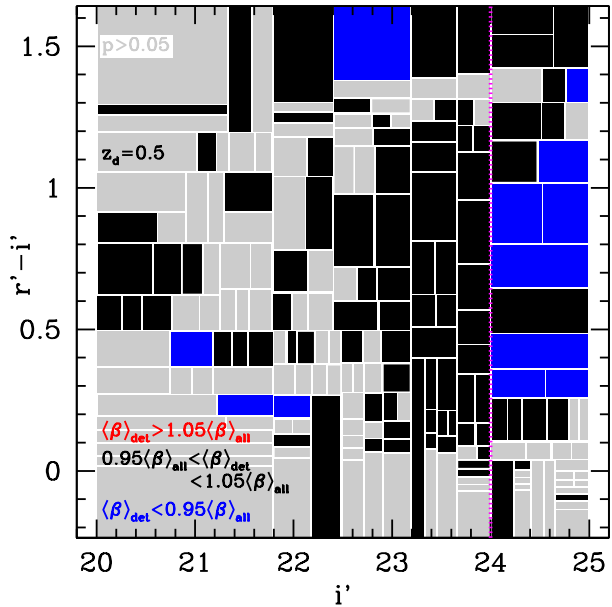} 
  \caption{CvM test of $p(z_{\rm phot})$ of all objects vs. objects detected in images to which noise was added such that the magnitude distribution $dN/di'$ peaks at $i'_t=24$ (indicated by magenta vertical, dashed line). Cells with significant differences in the distributions ($p<0.05$) are coloured according to the difference in mean $\beta$ between all objects and objects recovered from the noisy data according to their photometric redshift derived from the deep data.}
  \label{fig:detectioncvm}
\end{figure}
\begin{figure}
  \includegraphics[width=\linewidth]{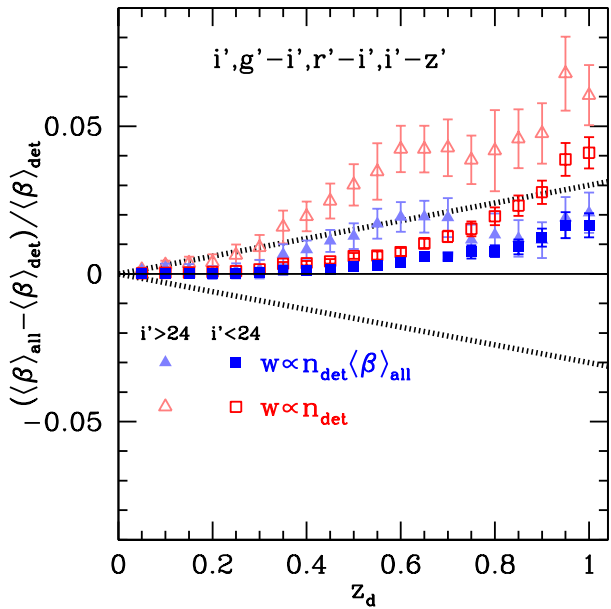} 
  \caption{Relative bias in mean $\beta$ (i.e. in estimated $\Delta\Sigma$) between magnitude-limited source sample at $20<i'<25$ and subsample detected in noisy data with peak of the magnitude distribution $dN/di'$ at $i'_t=24$. Blue filled symbols and red open symbols show minimum variance lensing weighting and uniform weighting of colour-magnitude cells in $g'r'i'z'$, respectively. Rectangular symbols show bright sample at $i'<24$, while faint, triangular symbols show cells with $i'>24$. The dotted, thick black lines indicate the approximate level of cosmic variance of lensing-weighted mean $\beta$ from the CFHTLS Deep reference catalogue (cf. \autoref{sec:cosmicvariance}).}
  \label{fig:detectionbias}
\end{figure}

In addition to random noise, other galaxy parameters (such as, e.g., size or morphology) might influence the detection probability of a source. If these parameters correlate with source redshift, the recovered source population might have a $\langle\beta\rangle$ that differs from the value of a truly magnitude limited sample to which the same colour-magnitude cuts have been applied. Because fields used for photometric measurements on reference samples of galaxies are (ideally) deeper, photometric selection effects will be weaker or nonexistent for the reference samples, potentially causing a bias.

We test this by adding noise to our CFHTLS Deep reference field images and re-running the photometric pipeline used to generate the magnitude limited reference catalogues. To this end, we use \textsc{SExtractor} background noise maps to determine the noise present in each pixel. We then add uncorrelated Gaussian noise to each pixel, such that the sum of the original variance and the variance of the artificial noise match a constant value for each of the CFHT pointings, chosen to have equal signal-to-noise in the aperture flux measurement of sources of the same extinction corrected flux in each of the pointings. We then re-run our object extraction, using a background noise map that is a constant at the respective value, masking a small number of pixels for which the original background noise already was above the target noise level. We chose the overall level of noise such that the magnitude histogram of the recovered sources peaks at $i'_t=24$ (cf. \autoref{fig:lf24}). In the resulting catalogue, the completeness is $\approx75\%$ at $i'_t=24$ and only $\approx20\%$ at $i'=25$, the completeness limit of the original data.

\autoref{fig:detectioncvm} shows the results of a CvM test of the $p(z)$ of the magnitude-limited vs. the detected subsample in $i',r'-i'$ space. Due to the much larger number of galaxies, small differences in the $p(z)$ of each colour-magnitude box can be detected with high significance. Such differences are present, especially at $i'>23$ (one magnitude brighter than the turnover of $dN/di'$). The difference of mean $\langle\beta\rangle$ is, however, far below the 5 per cent level in this regime. At the highly incomplete tail of $i'>24$, we find that the galaxies recovered from the noise tend to be at lower redshift, causing a considerable bias of $\beta$.

\autoref{fig:detectionbias} summarizes at the mean bias as a function of lens redshift for $g'r'i'z'$ information\footnote{The result is qualitatively independent of the combination of bands used}, split by source magnitude. Using only galaxies on the bright side of the turn-over point, $i'<24$, biases in optimally weighted source samples are always below the 2 per cent level. Biases in unweighted samples are larger, yet also at or below that level except for high lens redshift. Biases for the faint sample are only mildly worse in the weighted case, yet reach 5 per cent in the unweighted analysis at $z_d>0.5$. While there is a clear detection of selection bias, we note that cosmic variance from the limited number of four CFHTLS Deep reference fields (approximated by $\sigma_{\beta}/\beta\approx0.03 z_d$, cf. \autoref{sec:cosmicvariance}) is comparable to or larger than the bias caused by detection selection in the noisy data. The expected, even lower per cent level statistical uncertainty e.g.~for lensing mass calibration of optical cluster samples in final DES data, implies that both a treatment of this selection bias and an expansion of reference catalogues to reduce cosmic variance will be necessary.

\subsection{Shape measurement incompleteness and weighting}
\label{sec:shapebias}

Weak lensing shape catalogues of galaxies suffer another significant step of selection during star-galaxy separation and shape measurement. We investigate the effect of this on redshift biases by running a shape measurement pipeline on the CFHTLS Deep $i'$ band images and comparing the recovered to the full sample.

To this end, we perform a simple galaxy selection by a size cut above the stellar sequence \citep[cf., e.g.][their fig. 7]{2013MNRAS.tmp.1221G}, measure the shapes of galaxies with an implementation of the \citet*{1995ApJ...449..460K} algorithm, and apply cuts in shear responsivity, S/N of the shape measurement, and ellipticity \citep[as in][their section 4.2]{2014MNRAS.442.1507G}. 

\begin{figure}
  \includegraphics[width=\linewidth]{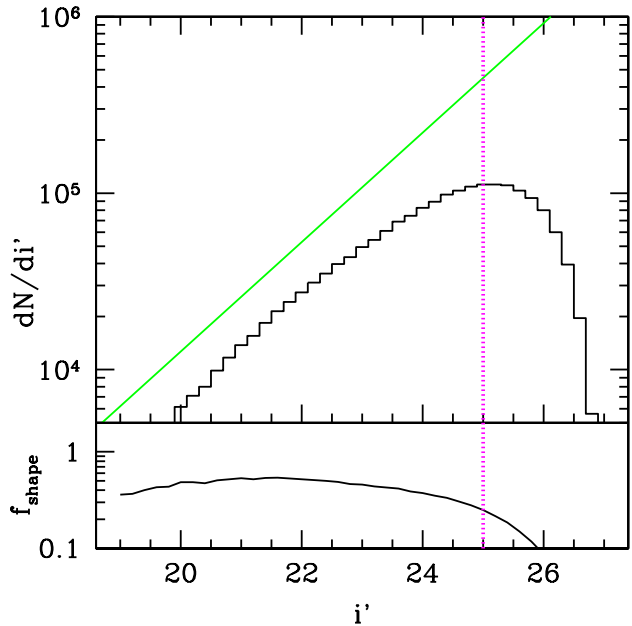}  
  \caption{Upper panel: Distribution of apparent \emph{i'} band magnitude of galaxies with successful shape measurement in the CFHTLS Deep fields. Green line indicates power-law fit to full photometric sample in the range $19<i'<22$ (cf.~\autoref{fig:lf}).  Lower panel: Completeness of shape catalogue relative to the power-law fit. Magenta lines indicates turnover magnitude $i'=25$, which is also the magnitude completeness limit of the photometric input catalogue. For comparison the figure is shown with the same axis scaling as \autoref{fig:lf}. }
\label{fig:lfshape}
\end{figure}

The resulting sample not only is somewhat shallower than the underlying magnitude-complete sample (see \autoref{fig:lfshape}), but also excludes a considerable fraction of brighter galaxies due to the size and responsivity cuts. We note that while the exact selection certainly differs between shape measurement pipelines, this qualitative observation generally holds \citep[cf., e.g.,][their Fig. 29]{2015arXiv150705603J}. The question is to what degree this morphology based selection changes the redshift distribution of the sample (see e.g. \citealt{2012arXiv1208.0605A}, who find biases at the few per cent level from approximating shape catalogue selection by a size cut). 

\begin{figure}
  \includegraphics[width=\linewidth]{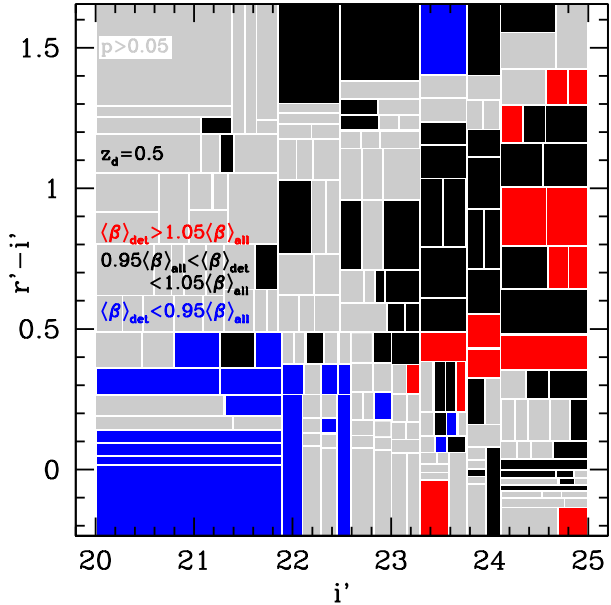}  
  \caption{CvM test of $p(z_{\rm phot})$ of all objects vs. objects with successful shape selection and measurement.  Cells with significant differences in the distributions ($p<0.05$) are coloured according to the difference in mean $\beta$ between all objects and objects recovered from the noisy data according to their photometric redshift derived from the deep data.}
\label{fig:shapecvm}
\end{figure}

\autoref{fig:shapecvm} shows the result of a CvM test of the compatibility of the $p(z_{\rm phot})$ of the magnitude complete sample and the shape catalogue subsample in cells in $i',r'-i'$. Significant differences in the redshift distributions are detected in large parts of colour-magnitude space. There appears to be a qualitative difference between the brighter subsample at $i'<23$, for which objects with successful shape measurement tend to be at lower redshift, and the faint subsample at $i'>23$, where the opposite is the case.

\begin{figure}
  \includegraphics[width=\linewidth]{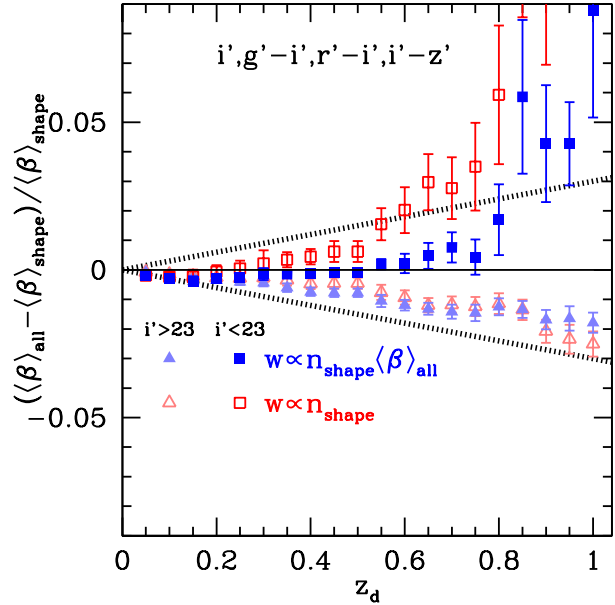}  
  \caption{Relative bias in mean $\beta$ (i.e. in estimated $\Delta\Sigma$) between magnitude-limited source sample at $20<i'<25$ and subsample selected for and successful in shape measurement. Blue filled symbols and red open symbols show minimum variance lensing weighting and uniform weighting of colour-magnitude cells in $g'r'i'z'$, respectively. Rectangular symbols show bright sample at $i'<23$, while faint, triangular symbols show cells with $i'>23$. The dotted, thick black lines indicate the approximate level of cosmic variance of lensing-weighted mean $\beta$ from the CFHTLS Deep reference catalogue (cf. \autoref{sec:cosmicvariance}).}
\label{fig:shapebias}
\end{figure}

For this reason, we investigate the impact of the shape selection on $\Delta\Sigma$ for the bright and faint samples separately in \autoref{fig:shapebias}. The bias of each sample is significant, but small compared to both the biases of spectroscopic selection described in \autoref{sec:specbias} and the cosmic variance from using only the four CFHTLS Deep fields for $p(z)$ estimation. This is especially true for the combination of bright and faint sample, in which the opposite effects partially cancel each other. At the 1-2 per cent level of relative error, shape selection remains a relevant factor, \red{especially with future surveys that intend to use larger reference catalogues for reducing cosmic variance}.

We also test the effect of weighting sources by their shape noise inverse variance, as it is commonly done to maximize S/N in weak lensing analyses. We find that the effect of this weighting on selection-based redshift biases is small compared to the above effect. Because the shape noise is strongly dominated by intrinsic dispersion of galaxy shapes in the case of CFHT Deep, where measurement noise is typically small, this conclusion could conceivably be different in more noisy data.

\section{Conclusions} \label{sec:conclusions}

We have developed a scheme for testing the impact of selection biases on empirical (training-based) $p(z)$ estimation methods for the purpose of weak lensing $\Delta\Sigma$ analyses. To this end, we use a reference catalogue with high-quality template-fitting photometric redshift point-estimates from CFHTLS Deep optical and WIRDS near-infrared broad-band photometry. To mimic empirical $p(z)$ estimation, we use a decision tree algorithm based on subsets of the available bands. We then apply this algorithm to estimate the $p(z)$ of galaxies in the reference catalogue itself as a target sample. In this, we impose selection effects and test their impact in terms of differences between the true and estimated distributions of $z_{\rm phot}$ of the target sample.

Selection effects in existing spectroscopic subsamples are significant over large parts of colour-magnitude space (cf. \autoref{sec:specbias}, \autoref{fig:cvm}). The dominant trend is that the existing spectroscopic subsamples are biased towards lower redshift than complete samples of equal magnitude and colour. Therefore the uniformly weighted mean $z$ and $D_{\rm ds}/D_{\rm s}$ estimated by empirical methods that rely on these spectroscopic training samples (Fig. 6) are also biased low, the latter at the 10 per cent level for moderate lens redshifts (\autoref{fig:betadiff}, open symbols). Because some parts of colour-magnitude space show opposite biases (Fig. 5), an optimally weighted $\Delta\Sigma$ measurement partially compensates the $D_{\rm ds}/D_{\rm s}$ offset (Fig. 7, filled symbols). \red{Because the redshift distributions of magnitude limited and spectroscopic subsamples at the same observed colour-magnitude differ, no re-weighting as a function of observed (photometric) galaxy properties can remedy these selection biases.}

Even with an optimal, magnitude-limited reference catalogue, selection biases in the lensing source catalogue due to object detection from noisy data and incomplete, morphology dependent success of shape measurement and star-galaxy separation can change the $p(z)$ distribution of the source sample. We test both of these effects (cf. \autoref{sec:photbias} and \ref{sec:shapebias}) and find them to be significant, although at a lower level than the aforementioned spectroscopic selection biases (cf. \autoref{fig:detectionbias} and \ref{fig:shapebias}). For the most common use cases ($\Delta\Sigma$ weighted measurements, lens redshifts below $z_d=0.8$) these selection biases are below the typical level of cosmic variance due to only having four square-degree reference fields of CFHTLS Deep data. \red{For future surveys that use larger reference catalogues for reducing cosmic variance, these effects may, however, be relevant.} In order to improve beyond 2 per cent level accuracy, empirical photo-$z$ methods will require both an increase in the number of reference fields and a treatment of selection effects in shape catalogues.

The dominant selection biases for empirical $p(z)$ estimation at this point therefore are due to target selection and incompleteness in the reference spectroscopic surveys. There are a number of potential avenues to address this issue:
\begin{itemize}
\item \citet{2015ApJ...813...53M} suggest the observation of representative spectroscopic samples over all relevant parts of colour-magnitude space, as defined by a self-organizing map (SOM). \emph{Representativeness} of the spectroscopic catalogue means two things: (1) sampling all of colour-magnitude space (as can be ensured by targeting galaxies in unpopulated cells of the SOM), and (2) ensuring that at a given colour-magnitude position there is no redshift dependence of spectroscopic incompleteness. We note that our tests of existing spectroscopic samples with claimed magnitude-limited target selection reveal non-negligible selection effects (cf. \autoref{fig:cvm}b), likely due to residual incompleteness of these surveys towards the faint end of their galaxy samples. Similar tests could be run on cells of the SOM.
\item Redshifts from angular cross-correlation of source galaxies with overlapping samples of know redshift do not require representative coverage with spectroscopic follow-up (see e.g.~\citealt{2006ApJ...651...14S,2008ApJ...684...88N,2013MNRAS.431.3307S,2013arXiv1303.4722M} and the recent application for cross-validation of the KiDS lensing $p(z)$ \citealt{2016arXiv160605338H,2016arXiv160909085M}). In addition, the method automatically accounts for selection effects in lensing source catalogues. These advantages make clustering redshifts a promising approach for future large-area surveys. Caveats include the requirement of moderately large samples of tracers with known $z$ out to large redshift, potential systematics from calibrating galaxy bias as a function of redshift \citep[e.g.][their section 3.1]{2015APh....63...81N}, and the need for a combined likelihood for cosmological parameters and $p(z)$ constrained from a joint data vector (including e.g.~lensing and galaxy clustering) and the measurements made for clustering redshifts.
\item A Bayesian scheme of fully representative galaxy templates and appropriate priors would allow for optimal $p(z)$ estimation from photometric data. While it is not evident how such a template set could be derived from a priori considerations or sparse spectroscopic surveys, a hierarchical Bayesian scheme such as that of \citet{2016MNRAS.460.4258L} could potentially derive coarse templates and priors  along with $p(z)$ from survey data \citep[see also][for estimating priors from the data]{2000ApJ...536..571B}. In order to break degeneracies of type and redshift in broad-band photometry, it might be necessary to include subsamples with narrow-band photometry or spectra in the hierarchical inference.
\item Surveys with numerous, narrow bands such as the Alhambra \citep{2008AJ....136.1325M}, PAU \citep{2014MNRAS.442...92M} and J-PAS \citep{2014arXiv1403.5237B} surveys, are in the process of deriving photometric redshifts with small uncertainties and outlier rates for large, magnitude-limited samples of galaxies. These samples might be well suited for empirical $p(z)$ estimation, although care must be taken to account for weak lensing source catalogue level selection at the few per cent level of systematic uncertainty.
\end{itemize}

There are thus multiple pathways to improve $p(z)$ estimation for the purpose of weak lensing. A framework like the one presented in this paper can provide a straightforward way of exploring those.

\section*{Acknowledgements}

This work is based on data made public by the CFHT Legacy Survey and the DEEP2, VIPERS, VUDS, VVDS-Deep and zCOSMOS spectroscopic surveys. The authors thank Steve Allen, Douglas Applegate, Nathalia Cibirka, Alexis Finoguenov, Enrique Gazta\~{n}aga, Will Hartley, Ben Hoyle, Ofer Lahav, Anja von der Linden, Peter Melchior, Anna Ogorzalek, Eduardo Rozo, Risa Wechsler, the anonymous referee and the DES and LSST photo-$z$ working groups for helpful discussions.

Support for DG was provided by NASA through the Einstein Fellowship Program, grant PF5-160138.

\appendix

\section{Statistical uncertainty of lensing reconstruction} 
\label{sec:variance}

In this appendix, we derive expressions for how the signal-to-noise ratio (S/N) of 
the lensing measurement is diminished when source galaxies are selected and weighted by their 
estimated redshift distribution rather than their true redshifts, using different combinations 
of filter bands. Results are given for deep ($i'<25$) and shallower ($i'<24$) samples for all
permutations of 2-4 band $g'r'i'z'$ photometry, in the hope that they might be useful in designing
photometric surveys.

For a set of sources $s=1,\ldots,N$ with true $\beta_s$ relative to some lens redshift $z_d$, 
the optimal estimator of gravitational shear combines the shears of all sources 
with a weight $w_s\propto\beta_s/\sigma_s^2$, where $\sigma_s$ contains intrinsic and measurement
related noise in the shape of source $s$. This relative weighting corresponds to the common
$\Delta\Sigma$ estimator (e.g. \citealt{Sheldon04.1}).

Let the tangential shear for source $s$ be $\gamma_s=\beta_s\gamma$, assume 
$\sigma_s=\sigma_{\gamma}$ is equal for all sources, and set $w_s=\beta_s/\sigma^2_{\gamma}$. 
The signal-to-noise ratio of the weighted estimate of $\gamma$ is 
\begin{equation}
\frac{\rm S}{\rm N}
=\frac{\sum\beta_s\gamma\times\beta_s/\sigma^2_{\gamma}}{\sqrt{\sum \sigma^2_{\gamma}\times\beta_s^2/\sigma^4_{\gamma}}}
=\frac{\gamma}{\sigma_{\gamma}}\sqrt{\sum\beta_s^2}
\end{equation}

If instead of the true value of $\beta_s$ we only have an estimate $\hat{\beta}_s$, the S/N becomes
\begin{equation}
\frac{\rm S}{\rm N}
=\frac{\sum\beta_s\gamma\times\hat{\beta}_s/\sigma^2_{\gamma}}{\sqrt{\sum \sigma^2_{\gamma}\times\hat{\beta}_s^2/\sigma^4_{\gamma}}}
=\frac{\gamma}{\sigma_{\gamma}}\frac{\sum\hat{\beta}_s\beta_s}{\sqrt{\sum \hat{\beta}_s^2}} \; .
\end{equation}

If the overall sample of $N$ galaxies is split into subsamples $j$ with $N_j$ galaxies each, 
and we have calibrated $\langle\beta\rangle_{s\in j}$ and assign it to each galaxy in the subsample (precisely what is done by our decision tree method), then
\begin{equation}
\frac{\rm S}{\rm N}
=\frac{\gamma}{\sigma_{\gamma}}\frac{\sum_j N_j \langle\beta\rangle_{s\in j}^2}{\sqrt{\sum_j N_j \langle\beta\rangle_{s\in j}^2}}
=\frac{\gamma}{\sigma_{\gamma}}\sqrt{\sum_j N_j \langle\beta\rangle_{s\in j}^2} \; .
\end{equation}

\begin{figure}
\centering
  \includegraphics[width=0.8\linewidth]{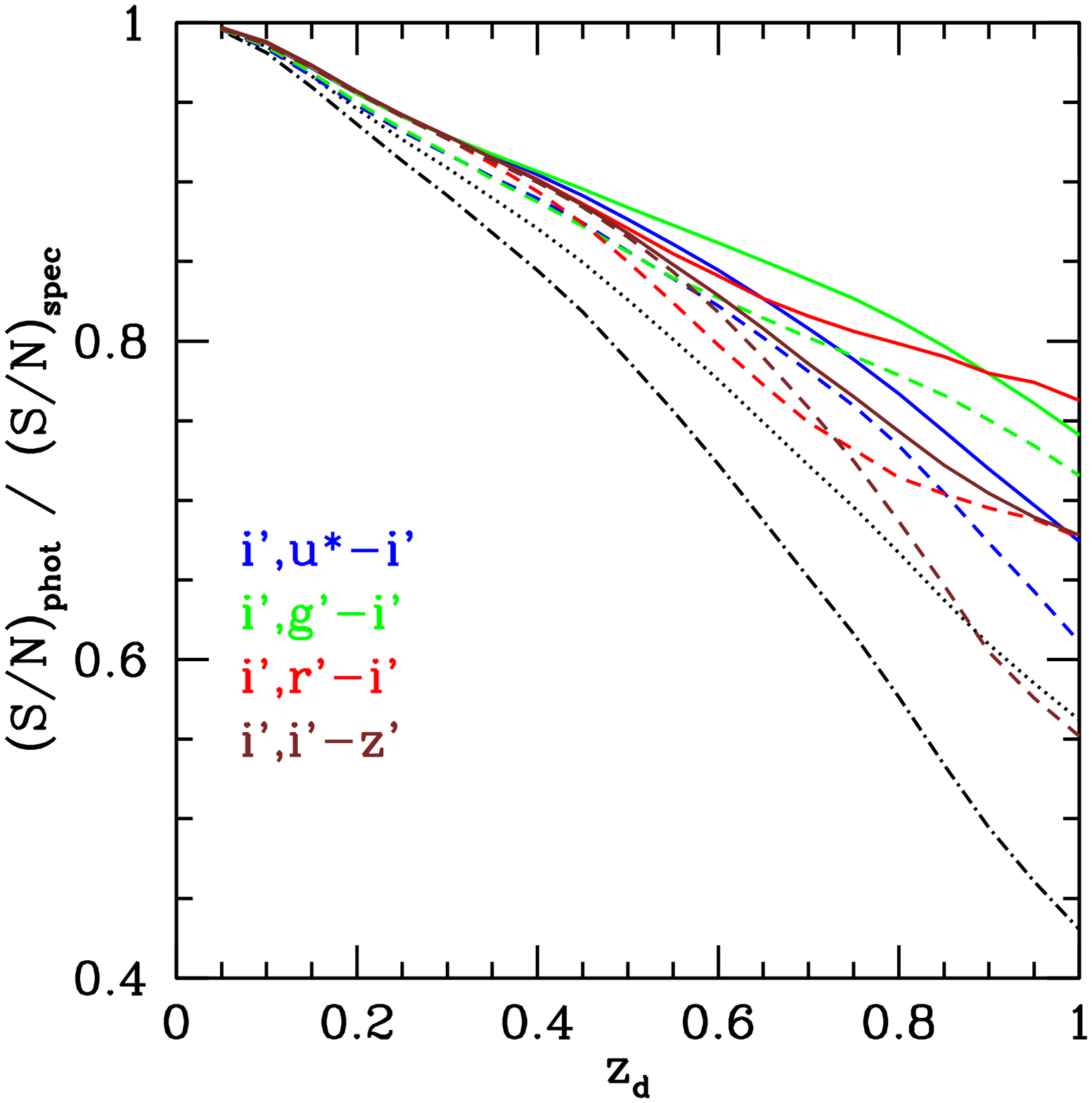}  
  \includegraphics[width=0.8\linewidth]{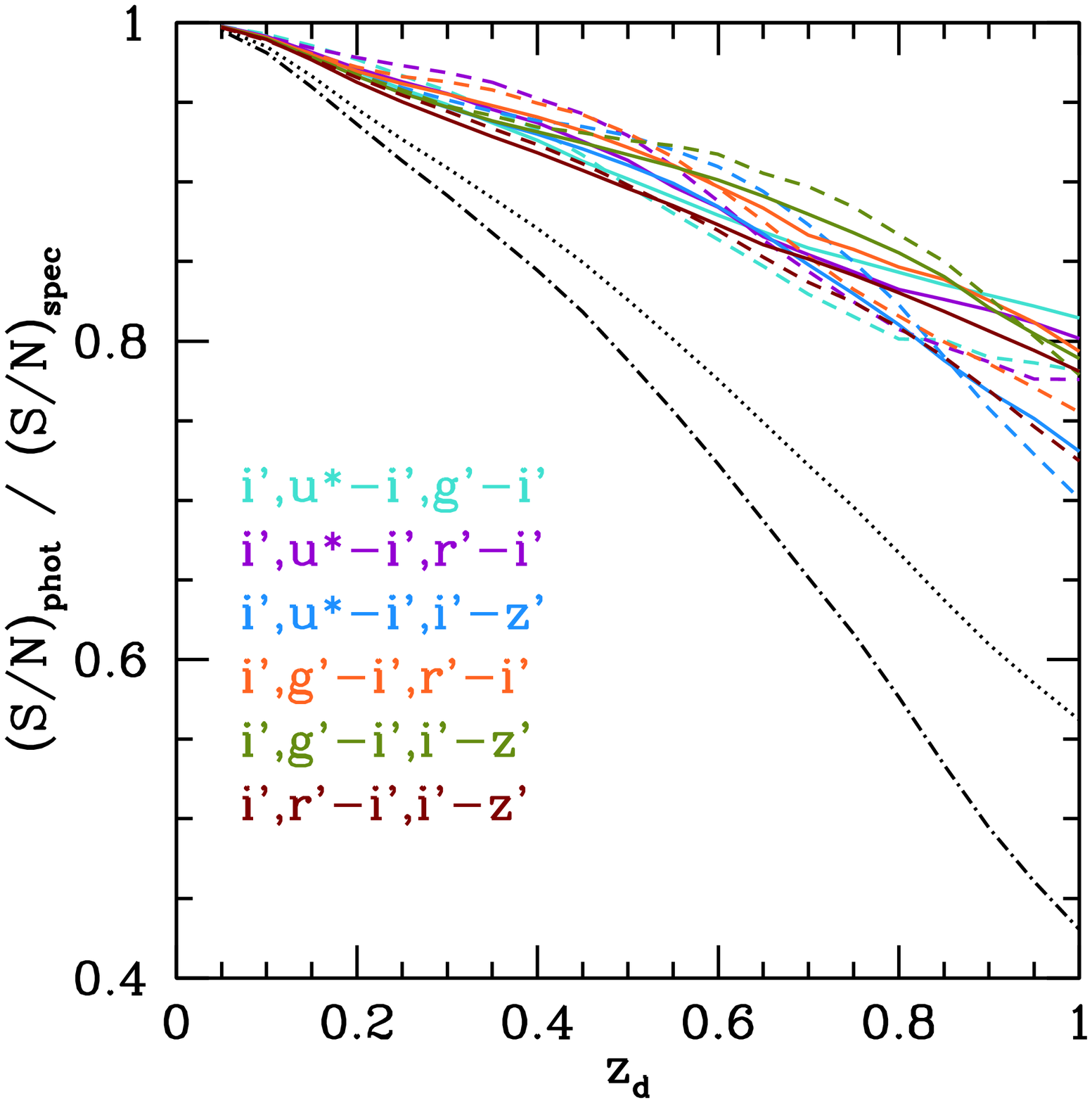}  
  \includegraphics[width=0.8\linewidth]{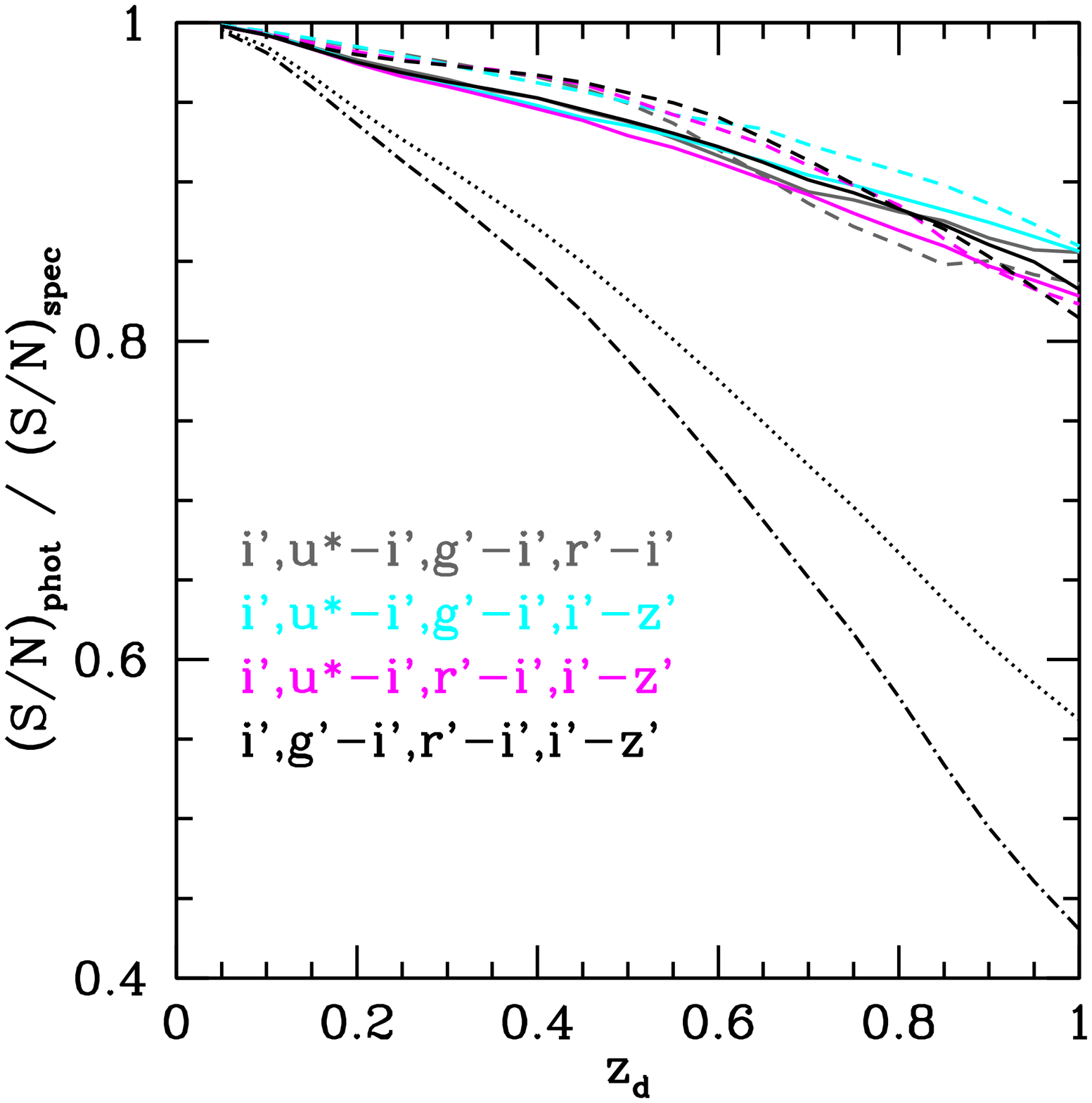}  
  \caption{Loss in lensing S/N relative to background selection and weighting from spectroscopic redshifts for $\beta$ estimated from a decision tree with 2, 3 or 4 band photometric information (top, central and bottom panel) as a function of lens redshift $z_d$. Solid (dashed) lines indicate the case of $20<i'<25$ ($20<i'<24$) source galaxies. Black dotted (dotted-dashed) lines show the loss in S/N in the case of assigning equal lensing weight to all sources.}
\label{fig:weightloss}
\end{figure}

For the simple case of a single sample with constant $\beta$ estimate, $\hat{\beta}_s=\langle\beta\rangle$, 
this reduces to
\begin{equation}
\frac{\rm S}{\rm N}
=\frac{\gamma}{\sigma_{\gamma}}\frac{\sum\langle\beta\rangle^2}{\sqrt{\sum\langle\beta\rangle^2}}
=\frac{\gamma}{\sigma_{\gamma}}\sqrt{\sum\langle\beta\rangle^2} \; .
\end{equation}

It therefore makes sense to define a metric that quantifies the fraction of the S/N ratio 
recovered when using $\hat{\beta}$ instead of $\beta$ as
\begin{equation}
\mathrm{rel.\,S/N}=\frac{\sqrt{\sum_j N_j \langle\beta\rangle_{s\in j}^2}}{\sqrt{\sum \beta_s^2}}=\sqrt{\frac{\langle\langle\beta\rangle_{s\in j}^2\rangle_j}{\langle\beta_s^2\rangle_s}} \; ,
\end{equation}
where the average over subsamples $\langle\ldots\rangle_j$ weights each subsample by its 
population $N_j$. 

\autoref{fig:tree} shows the relative S/N recovered in each colour-magnitude box based on the 
metric of the above equation by the colour saturation of the box lines drawn. \red{\autoref{fig:weightloss} shows the relative S/N of magnitude-limited samples from a colour-magnitude decision tree in all combinations of $i'$ and $u^{\star}g'r'z'$.}

While some combinations of two photometric bands do not recover much of the signal-to-noise lost by using the full sample with uniform weight (black lines), three-band (four-band) surveys consistently recover all but $30\times z_d$ per cent ($20\times z_d$ per cent) of the optimally weighted information. A similar statement can be made about tomographic cosmic shear analyses with colour-cut subsamples \citep{2007JCAP...03..013J}. We note, however, that also cosmic variance is a fairly strong function of the number and selection of bands used (see \autoref{sec:cosmicvariance}), and therefore samples selected with fewer bands require larger reference catalogues.

\section{Cosmic variance}
\label{sec:cosmicvariance}

Cosmic variance enters $\beta$ estimation from photometry because photometry does not
uniquely constrain a galaxy's redshift. The matter (and galaxy) density of a reference field
as a function of redshift therefore influences the estimated $p(z)$ of a given galaxy from
comparison to the reference sample. The level of cosmic variance, e.g. expressed in terms of the
relative systematic uncertainty of $\beta$, decreases with the number and area of reference fields used
and the discriminating power of the photometric information.

To determine the level of cosmic variance in our reference catalogues, we calculate Jackknife 
uncertainties for the estimated value of $\langle\beta\rangle$ for a lensing-weighted sample 
from splitting the reference catalogue into the four CFHTLS Deep fields. We build a decision 
tree as before, based on all sources. The value of $\langle\beta\rangle_j$ is calculated for 
each colour-magnitude box $j$ from the jackknifed subsample of reference objects. Color-magnitude 
boxes are weighted by the number and mean value of $\beta$ of objects in the full catalogue. 
We convert the jackknife uncertainty $\sigma_{\langle\beta\rangle}$ of the weighted mean $\langle\beta\rangle$ to a relative uncertainty $\sigma_{\langle\beta\rangle}/\langle\beta\rangle$.

\begin{figure}
  \includegraphics[width=\linewidth]{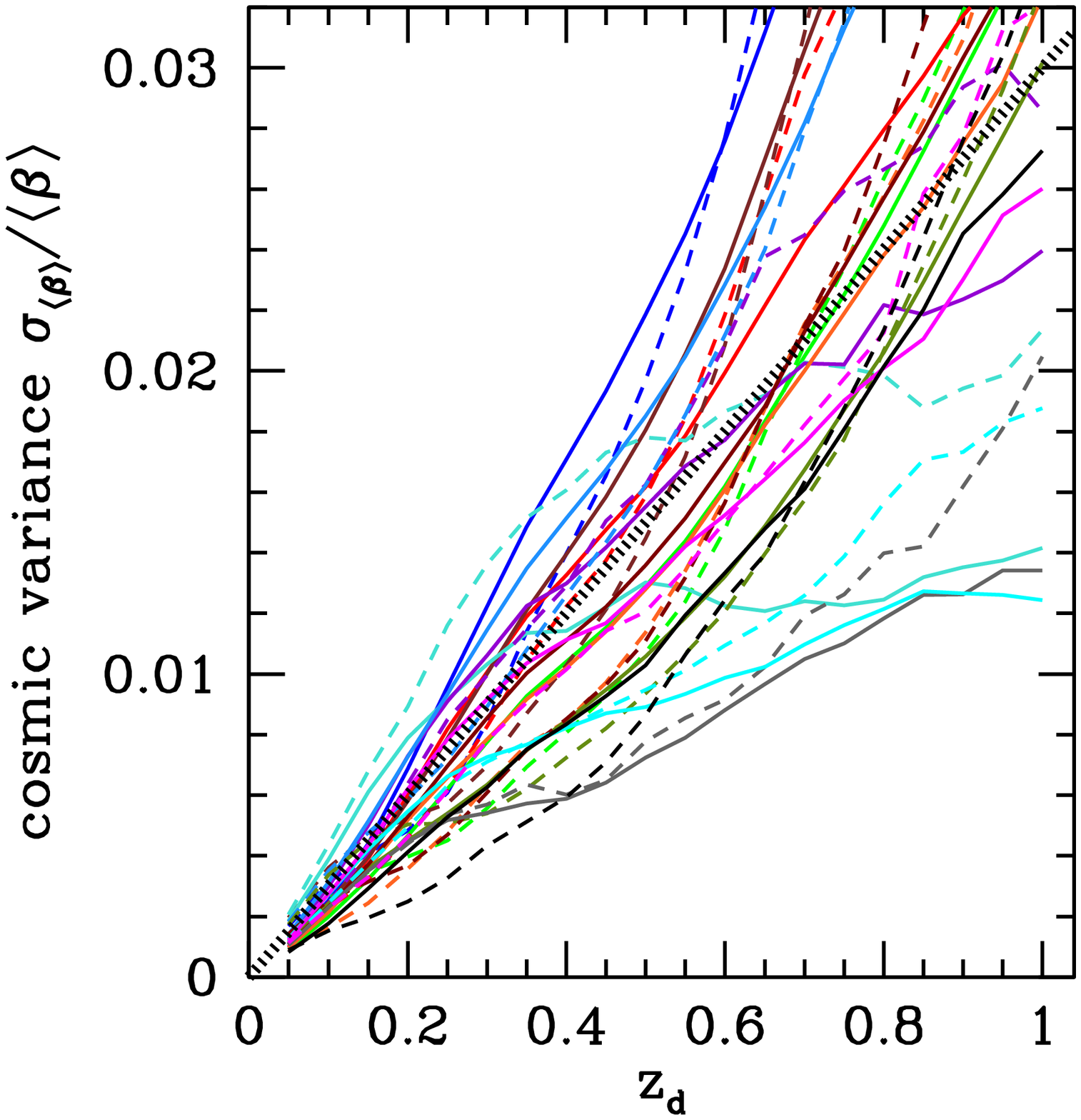} 
  \caption{Relative cosmic variance of the lensing-weighted value of $\beta$ of a sample 
           of galaxies with $20<i'<25$) (solid lines) and $20<i'<24$ (dashed lines) when 
           estimated from the four CFHTLS Deep fields. Color coding of the different 
           combinations of bands is as in \autoref{fig:weightloss}. Thick, dotted line 
           indicates $\sigma_{\langle\beta\rangle}/\langle\beta\rangle=0.03\times z_d$.}
  \label{fig:cv}
\end{figure}

Results are shown in \autoref{fig:cv} and indicate that, for a reference catalogue made of 
four independent square-degree pointings with the level of calibration \red{homogeneity} achieved for CFHTLS Deep, 
the relative level of cosmic variance is an approximately linear function of lens redshift, 
where suitable combinations of bands achieve $3\%\times z_d$ or better while less optimal
choices yields $\approx5\%\times z_d$. A larger number
of photometric bands typically allow for lower cosmic variance, down to $1.5\%\times z_d$. The most favorable 
combinations of bands generally include $u^{\star}$ band data. These results appear to be only a mild function of 
the depth of the sample.

For a reference catalogue compiled from $N_{\rm fields}$ independent square-degree fields, these uncertainties should scale as
$\sqrt{4/ N_{\rm fields}}$.

\section{Magnification bias}
\label{sec:magnification}

If a galaxy is \red{re-mapped by} a lensing Jacobian $A$, its surface area and observed flux increase by a magnification factor $\mu=\det A^{-1}$, which in the weak limit is approximated as $\mu\approx1+2\kappa$. At the same time, the surface number density of galaxies of these intrinsic properties decreases to $\mu^{-1}$ of its un-lensed value. Due to these two competing effects, the detected number density of galaxies in a magnitude limited sample can either increase or decrease. For a population with log-normal luminosity function $n(m)$ at the faint end with $\mathrm{d}\log n(<m)/\mathrm{d}m=s$, the change in number density is approximately $\Delta n/n=(5s-2)\kappa$ \citep[e.g.][]{1995ApJ...438...49B,2011ApJ...738...41U}.

Because $s$ depends on galaxy type and $\kappa\propto\beta$ depends on source redshift, magnification changes the colour-magnitude distribution and $p(z)$ of the observed galaxy population. In this Appendix, we study two aspects of that with the decision tree scheme:
\begin{itemize}
\item the difference between the estimated $\langle\beta\rangle_{\rm est}$ and the true $\langle\beta\rangle_{\rm true}$ of the magnified, magnitude limited population; this can be \red{thought} of as an effect of the prior in redshift, type and luminosity being wrong when estimated from the field population of a reference catalogue but applied to a magnified population
\item the difference between the estimated $\langle\beta\rangle_{\rm est}$ of a magnified population and the estimated field value $\langle\beta\rangle_{\rm field}$ of an unmagnified sample with the same magnitude limits; this is of importance when differences in estimated $p(z)$ between the field population and galaxy samples around clusters are used to estimate the contamination of the lensing source sample with cluster member galaxies (e.g.~for the purpose of estimating lensing boost factors, see \citealt{melchior}, their section 4.2)
\end{itemize}

To this end, we apply magnification to the reference catalogues and process them with a decision tree build on the un-magnified reference catalogue. Our methodological choices in this case are
\begin{itemize}
\item to consider a magnitude-limited sample (both in the magnified and unmagnified case) of $20<i'<24$, comparable to lensing source catalogues of ongoing large-area surveys;
\item to build a decision tree as described in \autoref{sec:build}, with a minimum number of 1000 reference sources per $g'r'i'z'$ colour-magnitude box;
\item to apply magnification due to a lens at some $z_d$ with $\kappa=0.1$ for sources at a hypothetical $\beta=1$; we ensured that the effect is well-described as linear in $\kappa(\beta)\beta^{-1}$ in the weak regime and describe the result as a relative change in $\langle\beta\rangle$ per unit $\kappa(\beta)\beta^{-1}$;
\item and to model magnification as a magnitude shift $\Delta m=-2.5\log(1+2\kappa)$, which moves some galaxies to different colour-magnitude boxes, and a rarefaction that corresponds to a relative weight of $w=(1+2\kappa)^{-1}$ of each galaxy
\end{itemize}

We estimate the respective versions of $\beta$ in a $\Delta\Sigma$ fashion, i.e.~weighting each colour-magnitude box $j$ by $N_j \langle\beta\rangle_{j, \rm field}$.

\begin{figure}
  \includegraphics[width=\linewidth]{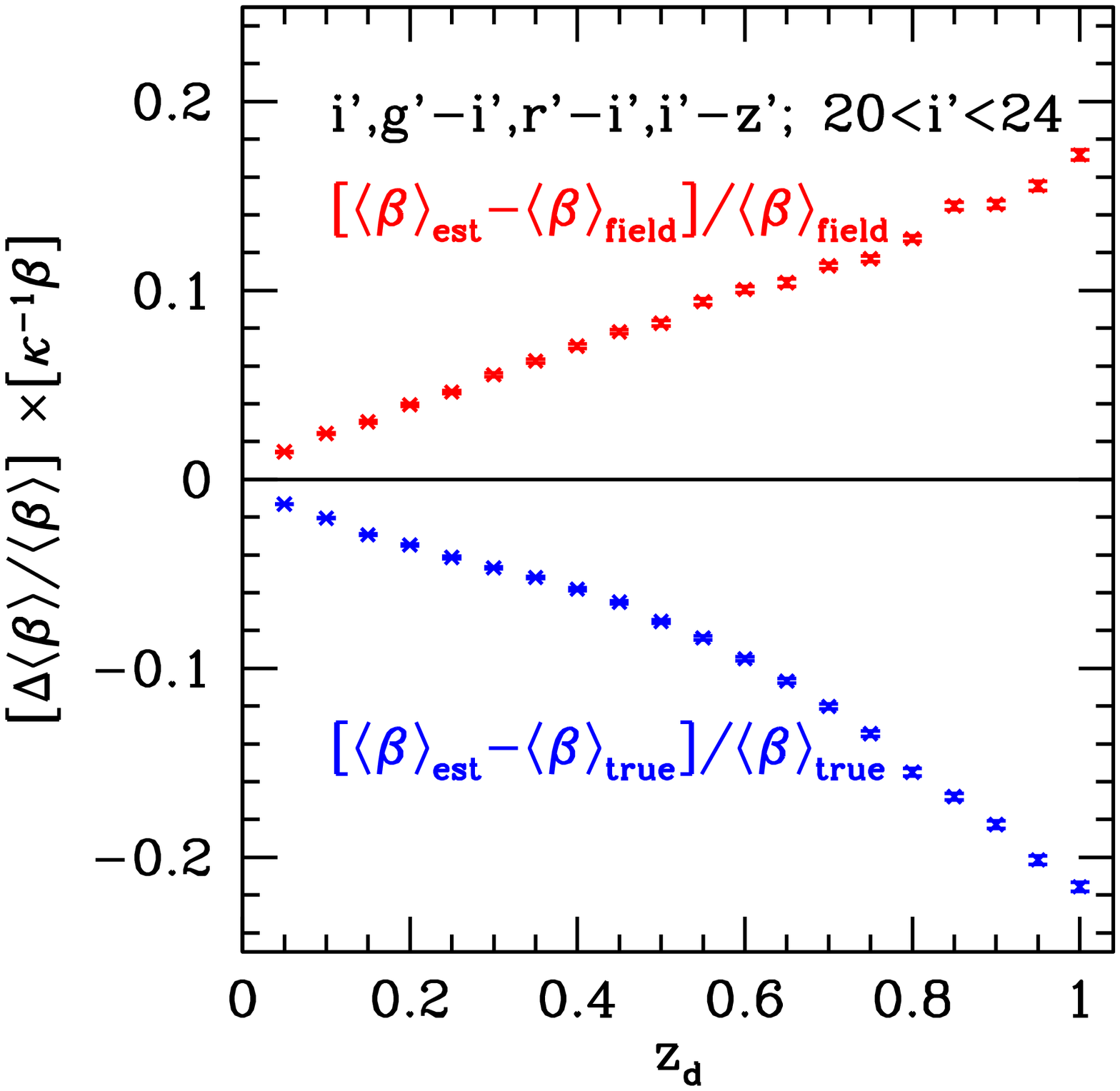} 
  \caption{Influence of lensing magnification of the background population on $\langle\beta\rangle$ estimates in a sample selected by observed magnitude $20<i'<24$. We show the relative bias due to mismatch between the field prior, $\langle\beta\rangle_{\rm est}-\langle\beta\rangle_{\rm true}$ of the magnified population (blue, always $<0$) and the \emph{estimated} relative difference  $\langle\beta\rangle_{\rm est}-\langle\beta\rangle_{\rm field}$ between the magnified and unmagnified population (red, always $>0$) as a function of lens redshift, for a hypothetical $\kappa=1$ at $\beta=1$.}
  \label{fig:magbias}
\end{figure}

We show the results of this test in \autoref{fig:magbias}. An example to read off from the plot is that for a lens at $z_d=0.6$ with a convergence of $0.1\times\beta$, the estimated $\langle\beta\rangle$ of a magnified background population with observed $20<i'<24$ is about 1 per cent smaller than the true $\langle\beta\rangle$, and about 1 per cent larger than the mean $\langle\beta\rangle$ of a field sample in this magnitude range.

These effects can affect cluster mass estimation with weak lensing in three ways:
\begin{itemize}
\item The underestimation of the $\beta$ of the magnified population causes an overestimate of mass, yet at an even smaller amplitude due to the much smaller convergence at larger radii from the cluster center, where most of the constraining power of the shear signal is located. 
\item The effect via boost factors estimated from $p(z)$ decomposition (\citealt{2014MNRAS.442.1507G}, their section 3.1.3 and \citealt[][their section 4.2]{melchior}) is a likely underestimate of background contamination (because the presence of cluster members should decrease the estimated $\langle\beta\rangle$, opposite to the effect of magnification) that causes an opposite bias of comparable amplitude.
\item For boost factors estimated from the number density of background sources, the effect will be opposite. For lenses at $0.4<z_d<1.0$, we find that the magnified sample in $20<i'<24$ has a higher surface density by $\approx\kappa\beta^{-1}\times10$per cent from magnification. This will bias boost factors high and, as a result, masses high.
\end{itemize}

\addcontentsline{toc}{chapter}{Bibliography}
\bibliographystyle{mn2e}
\bibliography{literature}

\label{lastpage}

\end{document}